\documentclass[aps,prl,twocolumn,superscriptaddress,showpacs,floatfix,letterpaper]{revtex4-1}%

\usepackage{graphicx}
\usepackage{tabularx}
\usepackage{bm}
\usepackage{latexsym}
\usepackage{epsf}
\usepackage{rotating}
\usepackage{epsfig,graphics,rotate,color}
\usepackage{wrapfig}
\usepackage{amssymb}
\usepackage{amsmath}
\usepackage{amsfonts}
\usepackage{subfigure}
\usepackage{array,hhline,dcolumn}%
\usepackage[normalem]{ulem}
\usepackage{color}
\usepackage[colorlinks=true,allcolors=blue]{hyperref}

\begin{document}
\bibliographystyle{apsrev4-1}

\title{Neutrino Physics Opportunities with the IsoDAR Source at Yemilab}

\author{J. Alonso}
\affiliation{
Massachusetts Institute of Technology, Cambridge, MA 02139, USA
}

\author{C.A.~Arg{\"u}elles}
\affiliation{Harvard University, Cambridge, MA 02138, USA}

\author{A. Bungau}
\affiliation{
Massachusetts Institute of Technology, Cambridge, MA 02139, USA
}

\author{J.M. Conrad}
\affiliation{
Massachusetts Institute of Technology, Cambridge, MA 02139, USA
}

\author{B. Dutta}
\affiliation{Texas A\&M University, College Station, TX 77840, USA}

\author{Y.D.~Kim}
\affiliation{
Center for Underground Physics, Institute for Basic Science (IBS), Daejeon 34126, Korea}
\author{E. Marzec}
\affiliation{
University of Michigan, Ann Arbor, MI 48109, USA
}
\author{D.~Mishins}
\affiliation{
University of Michigan, Ann Arbor, MI 48109, USA
}
\author{S.H. Seo}
\affiliation{
Center for Underground Physics, Institute for Basic Science (IBS), Daejeon 34126, Korea}
\author{M. Shaevitz}
\affiliation{
Columbia University, New York, NY 10027, USA
}

\author{J. Spitz}
\affiliation{
University of Michigan, Ann Arbor, MI 48109, USA
}

\author{A. Thompson}
\affiliation{Texas A\&M University, College Station, TX 77840, USA}

\author{L. Waites}
\affiliation{
Massachusetts Institute of Technology, Cambridge, MA 02139, USA
}

\author{D. Winklehner}
\affiliation{
Massachusetts Institute of Technology, Cambridge, MA 02139, USA
}

\begin{abstract}
IsoDAR seeks to place a high-power-cyclotron and target combination, as an intense source of $\bar{\nu}_e$ at the level of $\sim 10^{23}$/year, close to a kiloton-scale neutrino detector in order to gain sensitivity to very short-baseline neutrino oscillations ($\bar{\nu}_e \rightarrow \bar{\nu}_{e}$) and perform precision tests of the weak interaction, among other physics opportunities. Recently, IsoDAR has received preliminary approval to be paired with the 2.26~kton target volume liquid scintillator detector at the Yemi Underground Laboratory (Yemilab) in Korea, at a 17~m center-to-center baseline, and cavern excavation for IsoDAR is now complete.   In this paper, we present the physics capabilities of IsoDAR@Yemilab in terms of sensitivity to oscillations (via inverse beta decay, IBD; $\bar{\nu}_e+p \rightarrow e^+ + n$), including initial-state wavepacket effects, and the weak mixing angle (via elastic scattering off atomic electrons, $\bar{\nu}_e + e^- \rightarrow \bar{\nu}_e + e^-$).  We also introduce a study of IsoDAR sensitivity to new particles, such as a light $X$ boson, produced in the target that decays to $\nu_e \bar \nu_e$.
\end{abstract}
\maketitle

\section{Introduction}
The IsoDAR concept, in which a powerful and compact cyclotron is brought close to a large existing or planned underground detector, represents a significant paradigm shift in neutrino physics. Such an experiment would open the possibility for new physics discoveries in various forms, including neutrino production, interactions, and oscillations, each of which would present as unexpected spectral deviations in the high statistics event samples observed at the detector. In addition to the particle physics opportunities enabled by IsoDAR and as detailed in a number of publications, including outside of neutrino physics~\cite{Hostert:2022ntu}, the experiment will be especially important for applications in accelerator and medical science as well~\cite{accelerator,medical}. 

\begin{figure}[h]
\begin{centering}
\includegraphics[width=8.7cm]{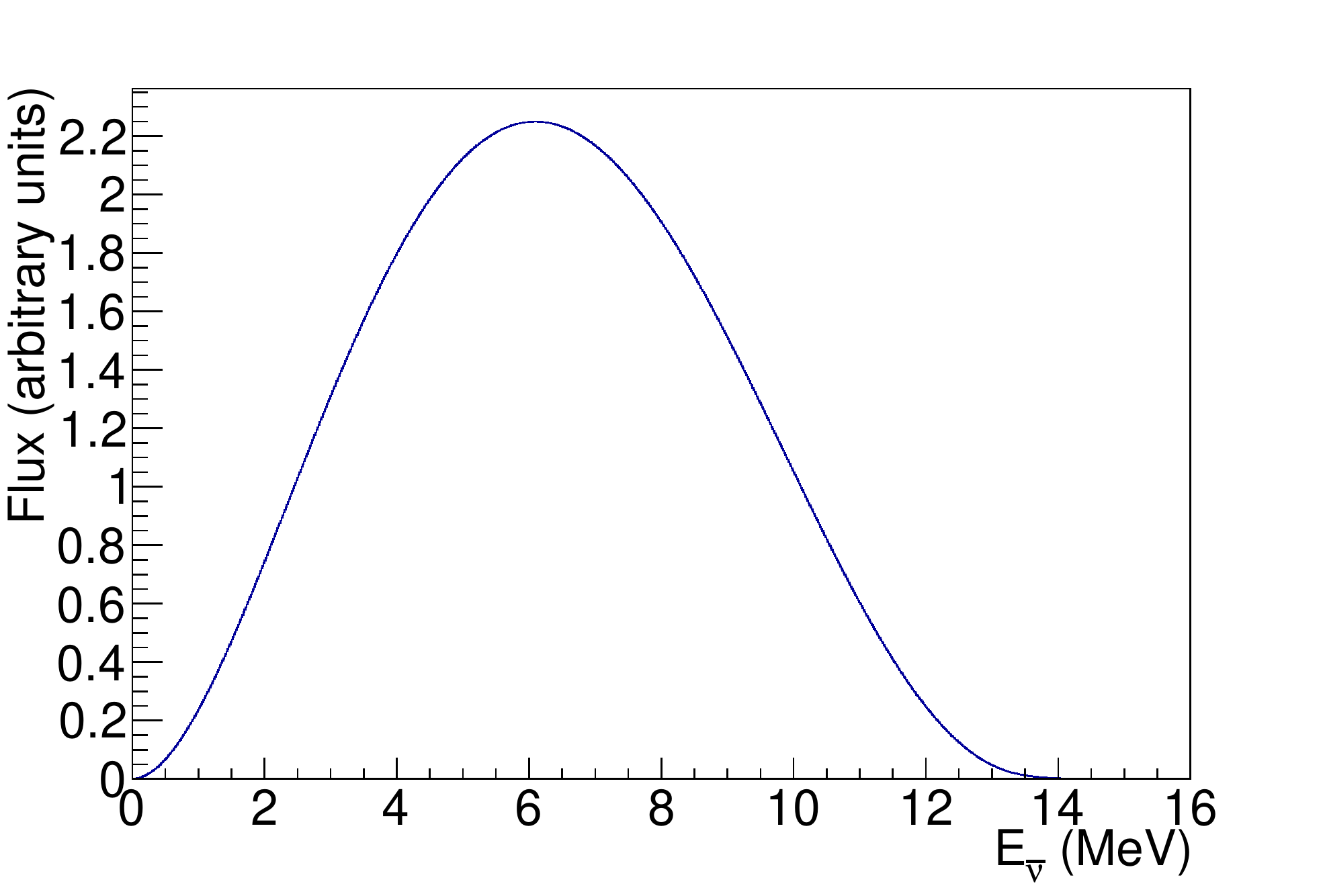}
\vspace{-.6cm}
\caption{The IsoDAR $\bar{\nu}_e$ flux arising from $^8$Li beta decay, adapted from Ref.~\cite{isodar_prl}.}
\label{flux}
\end{centering}
\end{figure}

IsoDAR will rely on 60~MeV proton interactions with a $^9$Be target (600~kW) to produce a powerful source of neutrons. Neutron capture on the surrounding $\geq$99.99\% isotopically pure $^7$Li sleeve results in an intense source of $\bar{\nu}_e$ from the high-$Q$ $\beta$-decay of $^8$Li ($\rightarrow~^8\mathrm{Be}+e^-+\bar{\nu}_e$; $\tau_{1/2}=839$~ms) with a mean antineutrino energy of 6.4~MeV and an endpoint of $\sim$15~MeV. With $1.97 \cdot 10^{24}$ protons on target per year and 0.015~$\bar{\nu}_e$/proton, IsoDAR will produce $1.15 \cdot 10^{23}$ $\bar{\nu}_e$ in 4~years of livetime (5 years of running at 80\,\% duty cycle); the $\bar{\nu}_e$ flux shape is shown in Figure~\ref{flux}.

 In this paper, we consider the physics capabilities of the IsoDAR $\bar{\nu}_e$ source (accelerator+target) paired to a planned 2.26~kton (inner volume) detector, called the Liquid Scintillator Counter (LSC), at the Yemilab Center for Underground Physics in Korea at a center-to-center distance of 17~m~\cite{isodar_yemilab,Alonso:2022mup}.   
The envisioned detector is cylindrical with 7.5~m radius and 15~m height (inner volume), along with a 1~m buffer region extending from the top and sides, and a 1.5~m veto region extending further. The liquid scintillator properties and photocoverage are expected to be similar to KamLAND~\cite{KamLAND:2013rgu}. In addition, the radiopurity capabilities of the detector, which are expected to surpass KamLAND, and prospects of reconstructing $e^\pm$ direction, are discussed below.  This allows for expanded physics capability beyond the pairing of IsoDAR at KamLAND that has been the focus of previous publications ~\cite{isodar_prl, toups}.    In this paper, we describe the improved sensitivity for neutrino oscillation and electroweak measurements, as well as introduce an additional physics goal: the search for unexpected peaks in the $\bar{\nu}_e$ flux due to novel physics, such as a new light boson.   The excavation for IsoDAR rooms was completed in January 2022 and construction of the LSC hall is well underway. 

The possibility of new physics associated with neutrino mixing at short-baselines has, simultaneously, never been stronger \textit{and} more confusing. The MiniBooNE experiment's latest result~\cite{miniboone_new} shows a 4.8$\sigma$ indication of $\nu_\mu \rightarrow \nu_e$ oscillations with a characteristic $\Delta m^2\sim1~\mathrm{eV}^2$, consistent with MiniBooNE's 2.8$\sigma$ evidence of $\bar{\nu}_\mu \rightarrow \bar{\nu}_e$~\cite{MB_antinu} and LSND's 3.8$\sigma$ evidence of $\bar{\nu}_\mu \rightarrow \bar{\nu}_e$~\cite{lsnd, lsnd2, lsnd3}. The compatibility of these results with each other and seemingly with the reactor-~\cite{reactor} and radioactive-source-based~\cite{source,2109.11482} electron-flavor disappearance anomalies, perhaps within a 3+1 model with mixing among the three active flavors and one ``sterile" flavor, is generally stymied by the global lack of observed muon-flavor disappearance, although a notable exception to this comes from IceCube~\cite{IceCube} (see discussion below). The varied results may be indicative of a more complicated modification to 3 neutrino mixing beyond this simple extension.   This paper updates a 3+2 scenario we have previously presented, and adds a study of a recently proposed scenario involving sterile neutrino decay~\cite{decay1, decay2, decay3, decay4, decay5, decay6, decay7, decay8, decay9, decay10, decay11} motivated by new results from IceCube~\cite{Moulai:2021zey}. 
IsoDAR@Yemilab will provide unsurpassed sensitivity to $\bar{\nu}_e \rightarrow \bar{\nu}_{e}$ oscillations observed as a periodic deficit in the well-predicted IBD signal
due to the increased size of the detector and longer $L$.  The ability to trace an $L/E$-dependent wave over many cycles (in much of the currently-favored parameter space), and/or more unusual $L$- or $E$-dependent behavior, is unique to IsoDAR and is likely to disentagle this complicated situation.

In addition to measuring IBD events in the context of an oscillation search, IsoDAR@Yemilab can use this large sample of events for a generic (or model-dependent) ``bump hunt'' --- a search for a peak in the well predicted IBD event rate versus energy. Such a search is well motivated by theoretical interest in light-mass mediators and a number of experimental anomalies, including the ``5~MeV bump''~\cite{prospect2,stereo,NEOS:2016wee,reno,dayabay,doublechooz} observed in numerous reactor experiments and a possible ``X17" particle~\cite{Krasznahorkay:2015iga}.

Similar to precision oscillation studies and an IBD-based ``bump hunt", neutrino-based measurements of the weak mixing angle, $\sin^2{\theta_W}$, are also highly sensitive to new physics. Specifically, non-standard neutrino interactions (NSI) can present as a deviation from the well-predicted $\sin^2{\theta_W}$-dependent Standard Model (SM) cross section of $\bar{\nu}_e + e^- \rightarrow \bar{\nu}_e + e^-$. The prospect of improved electroweak measurements using neutrinos is especially exciting because the most precise measurement to date with neutrinos, coming from the NuTEV experiment~\cite{nutev}, deviates from the SM prediction, constrained by the electroweak measurements from LEP~\cite{LEP}, by $\sim3\sigma$. While a number of possible explanations for this anomaly exist, often involving modified nuclear physics assumptions required to extract the neutral-current to charged-current and neutrino-nucleus to antineutrino-nucleus cross section ratios relevant for NuTEV's measurement, a definitive explanation of this long-standing anomaly remains elusive. Either way, it is clear that improved neutrino-based $\sin^2{\theta_W}$ measurements across many energy scales are valuable, noting that NuTEV's measurement was at $\mu\sim4.5$~GeV. At the MeV-scale energies relevant for this discussion, global measurements of $\sin^2{\theta_W}$ are extremely sparse with only $\sim$1000 total $\bar{\nu}_e-e^-$ events collected at reactors ($\sim$1-10~MeV)~\cite{irvine,texono,rovno,munu} and accelerators ($\sim$10-50~MeV)~\cite{lampf,lsnd}. IsoDAR@Yemilab will collect about a factor of 7 more than this worldwide data sample.

This paper is organized as follows: first, we consider the sensitivity of ``IsoDAR@Yemilab" (17~m center-to-center from IsoDAR to the Yemilab detector) to $\bar{\nu}_e \rightarrow \bar{\nu}_{e}$ mixing via IBD detection; next, we provide an in-depth discussion of this oscillation sensitivity in the context of global searches for short-baseline oscillations, including mixing beyond the simplest 3+1 model; next, we consider the implications of decoherence on IsoDAR@Yemilab's sensitivity to oscillations, including prospects for measuring the finite initial antineutrino wavepacket; next, we discuss the prospects of a ``bump hunt" in the measured IBD spectrum, especially in the context of a search for a new boson; finally, we study IsoDAR@Yemilab's ability to measure $\sin^2{\theta_W}$ and search for non-standard interactions (NSI) via $\bar{\nu}_e$ elastic scattering off atomic electrons (ES; $\bar{\nu}_e + e^- \rightarrow \bar{\nu}_e + e^-$). 

\begin{figure}[h]
\begin{centering}
\centering
\includegraphics[width=8.5cm]{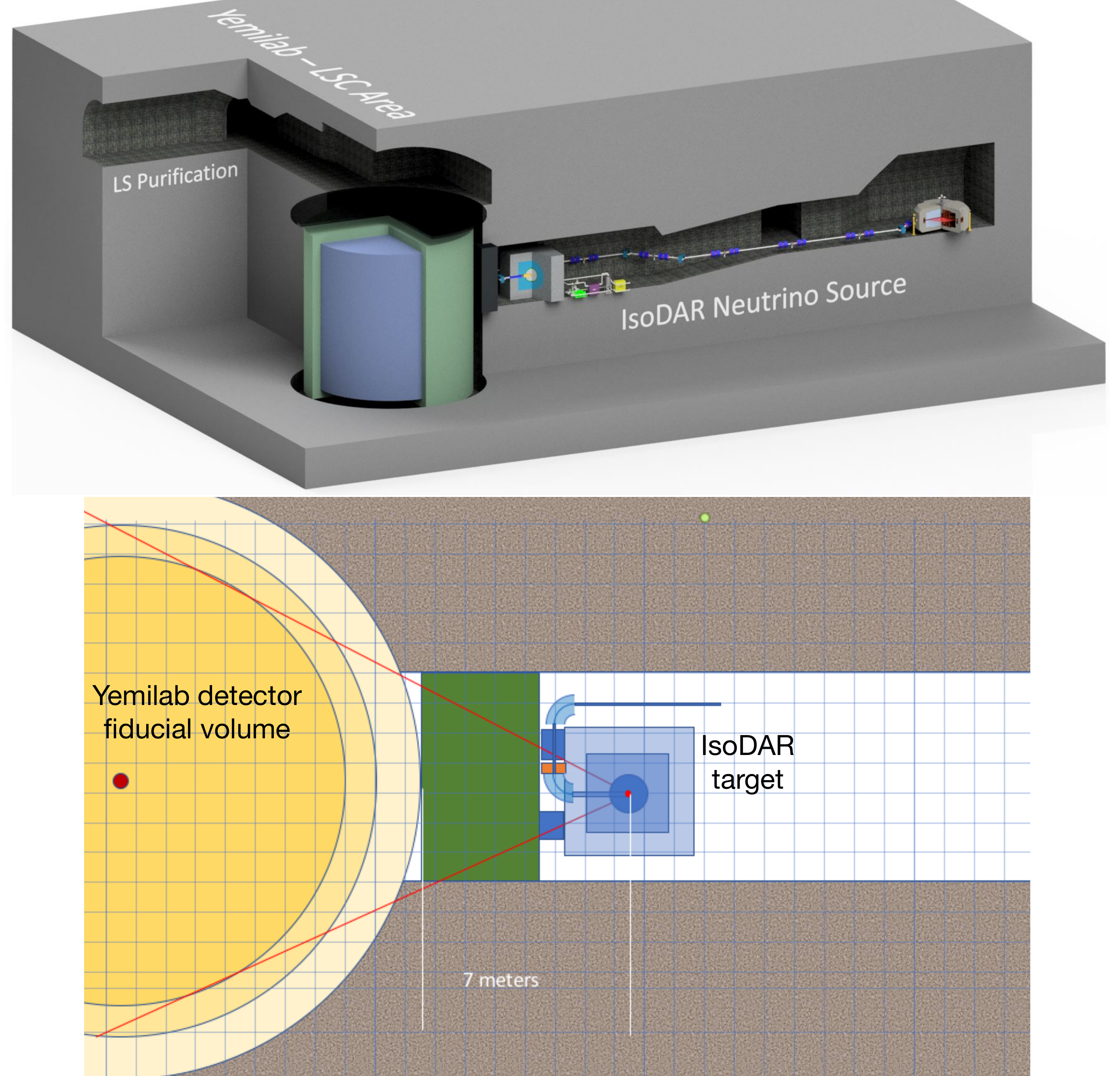}
\caption{The IsoDAR@Yemilab configuration and geometry considered for the studies here. The distance from the center of the IsoDAR target to the center of the detector is 17~m (from left-to-right in the bottom drawing, 7.5~m inner detector radius, 1.0~m buffer, 1.5~m veto, and 7~m of shielding+target).}
\label{yemilab_geometry}
\end{centering}
\end{figure}

For the IsoDAR@Yemilab scenario considered, we place the IsoDAR target center at the mid-plane of the detector, next to the outer-tank. The distance from the center of the IsoDAR target to the center of the Yemilab detector in this configuration is 17~m: 7.5~m (radius of the detector) + 1~m (detector buffer region) + 1.5~m (detector veto region) + 7~m (shielding and IsoDAR beam-pipe/target geometry). The envisioned source and detector geometry is shown in Fig.~\ref{yemilab_geometry}. While this conceptual arrangement may require adjustments based on ongoing engineering and shielding studies, we expect it to be representative of what is achievable for this ``IsoDAR@Yemilab" experiment. For reference, the relevant IsoDAR accelerator/target and detector assumptions applicable to all analyses presented are shown in Table~\ref{assumptions_table_1}. The IsoDAR-specific parameters are based on Ref.~\cite{accelerator}. Notably, the IBD-based and ES-based analyses feature significantly different signal and background event rates, and thus require separate fiducial volume definitions towards optimizing the sensitivity of each. The analysis-specific assumptions are discussed below. 

\begin{table}[h]
      \begin{tabular}{|c|c|} \hline 
Runtime  &  5 calendar years  \\ \hline
IsoDAR duty factor  &  80\%  \\ \hline
Livetime  &  4 years  \\ \hline
Protons on target/year  &  $1.97\cdot 10^{24}$  \\ \hline
$^8$Li/proton ($\bar{\nu}_e$/proton) &  0.0146  \\ \hline
$\bar{\nu}_e$ in 4 years livetime  &  $1.15\cdot 10^{23}$  \\ \hline
IsoDAR@Yemilab mid-baseline  &   17~m  \\ \hline
IsoDAR@Yemilab depth  & 985~m (2700 m.w.e.) \\ \hline
\end{tabular}
\caption{Accelerator/target assumptions and detector specifications for the IsoDAR@Yemilab experiment.}
\label{assumptions_table_1}
\end{table}

\section{Sensitivity to electron antineutrino disappearance \label{nuedis}}

The IsoDAR@Yemilab experiment will be able to collect $1.67\cdot 10^{6}$ IBD events in 5~years of running. 
In addition, the wide range of baselines (9.5-25.6~m) and $\bar{\nu}_e$ energies (0-15~MeV) in this setup combined with the reconstruction abilities of the detector affords strong sensitivity to $\bar{\nu}_e \rightarrow \bar{\nu}_{e}$ disappearance as a function of $L/E$ (and, $L$ or $E$ considered individually). For detecting IBD events in the Yemilab detector, our assumptions are shown in Table~\ref{assumptions_table_2}. 


\begin{table*}[t]
\textbf{IBD analysis assumptions}  \\ 
      \begin{tabular}{|c|c|} \hline 
IsoDAR@Yemilab baseline range  &   9.5-25.6~m \\ \hline 
IsoDAR@Yemilab fiducial mass  &  2.26 kton  \\ \hline
IsoDAR@Yemilab fiducial size (radius, height)  &  7.5~m, 15.0~m \\ \hline
1$\sigma$ uncertainty in $\bar{\nu}_e$ creation point  &  0.41~m  \\ \hline
Prompt (e$^+$) energy res.   &  $\sigma(E)=6.4\%/\sqrt{E~\mathrm{(MeV)}}$  \\ \hline
Prompt (e$^+$) energy res. @ 8~MeV  &  2.3\%  \\ \hline
Prompt (e$^+$) vertex res.   &  $\sigma[\mathrm{vertex~(cm)}]=12/\sqrt{E~\mathrm{(MeV)}}$ \\ \hline
Prompt (e$^+$) vertex res. @ 8~MeV  &  4~cm \\ \hline
Total $\bar{\nu}_e$ IBD efficiency  &  92\%   \\ \hline \hline
Total detected $\bar{\nu}_e$ IBD (92\% efficiency) &  1.67$\cdot 10^6$  \\ \hline
\end{tabular}
\caption{The assumptions relevant for the IsoDAR@Yemilab IBD-based analyses.}
\label{assumptions_table_2}
\end{table*}

\begin{figure}[h]
\begin{centering}
\includegraphics[width=8.5cm]{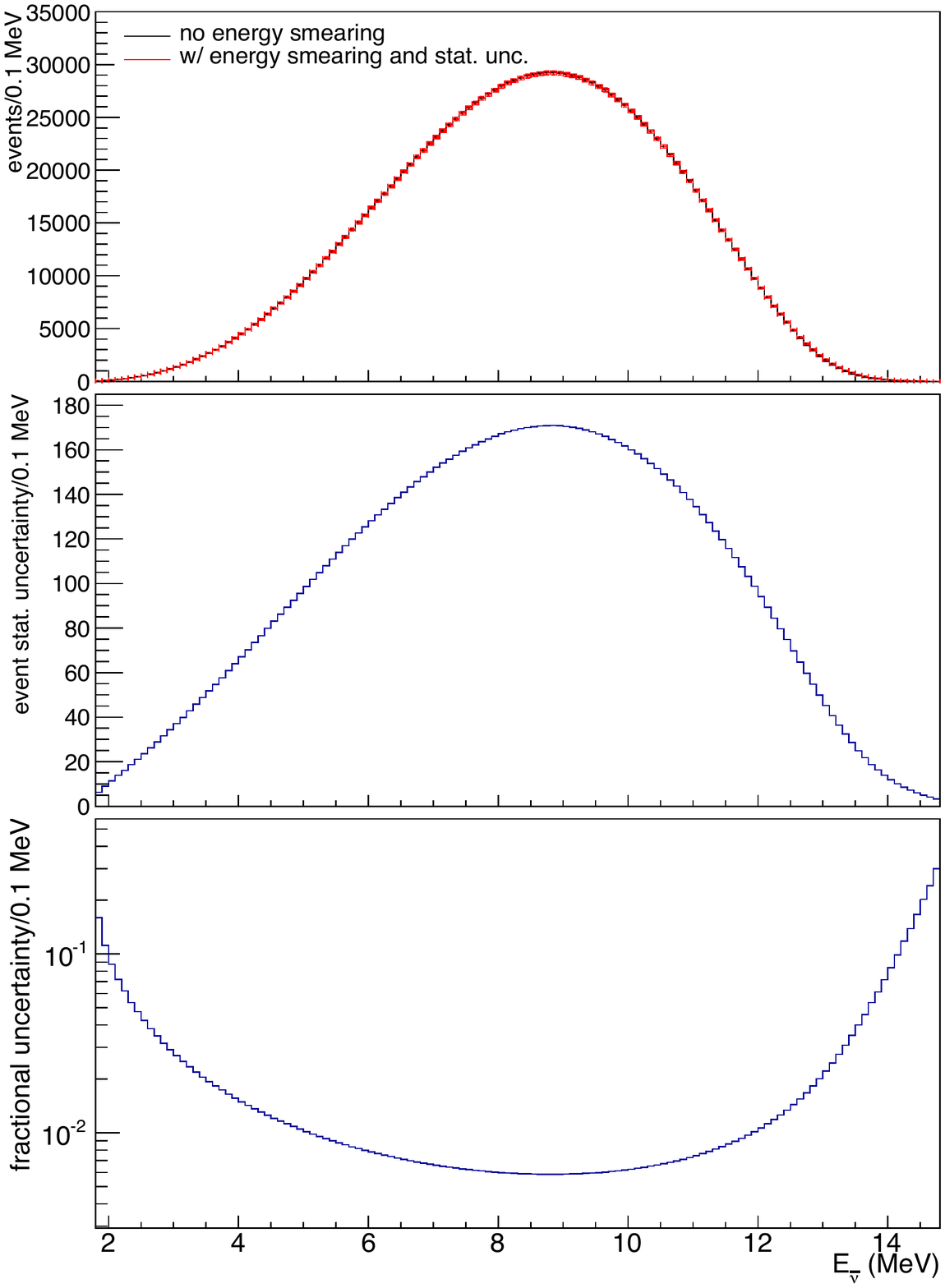}
\vspace{-.6cm}
\caption{Top: the total IBD rate versus antineutrino energy for 4~years livetime with IsoDAR@Yemilab, showing both true and reconstructed energy and statistical error bars on the reconstructed energy distribution given a bin width of 0.1~MeV. Notably, the effect of energy smearing is very hard to discern even at this small bin width. Middle (Bottom):  Absolute (fractional) statistical uncertainty in number of events as a function of reconstructed antineutrino energy.}
\label{ratesforIBD}
\end{centering}
\end{figure}

For IBD events, the $\bar{\nu}_e$ energy is reconstructed based on the positron energy, $E_{\bar{\nu}_e}=E_{e^+}+0.78$~MeV, and the IBD detection efficiency at the Yemilab detector is assumed to be $92\pm0.7$\,\%, consistent with KamLAND~\cite{KamLAND:2010fvi,Gando}. Given the $L/E$-dependent behavior of conventional neutrino oscillations, the energy and vertex (or, baseline) resolutions are highly relevant for an oscillation search. The energy and vertex resolutions are assumed to be consistent with KamLAND, $\sigma(E)=6.4\%/\sqrt{E~\mathrm{(MeV)}}$ and $\sigma[\mathrm{vertex~(cm)}]=12/\sqrt{E~\mathrm{(MeV)}}$~\cite{KamLAND:2010fvi}.   Fig.~\ref{ratesforIBD} (top) compares the true IBD event energy spectrum in black to the reconstructed IBD energy spectrum in red, and one sees that the reconstructed energy is highly precise and, even with only 0.1 MeV bin width, the uncertainties are small. The middle and bottom panels of Fig.~\ref{ratesforIBD} show the absolute and fractional expected statistical uncertainty as a function of the reconstructed antineutrino energy.

With regard to vertex resolution, however, we note that the inherent uncertainty in the $\bar{\nu}_e$ baseline, essential for oscillation studies, on an event-by-event basis is dominated by knowledge of the $\bar{\nu}_e$ creation position in the IsoDAR target and sleeve. Based on the currently envisioned IsoDAR target/sleeve geometry~\cite{Bungau:2018spu} and \textsc{Geant4}-based simulations~\cite{GEANT4:2002zbu}, the characteristic uncertainty in the creation position is modeled with a 41~cm 1$\sigma$ spherical Gaussian uncertainty.  

\begin{figure*}[t!]       
\begin{center}
\includegraphics[width=7.1in]{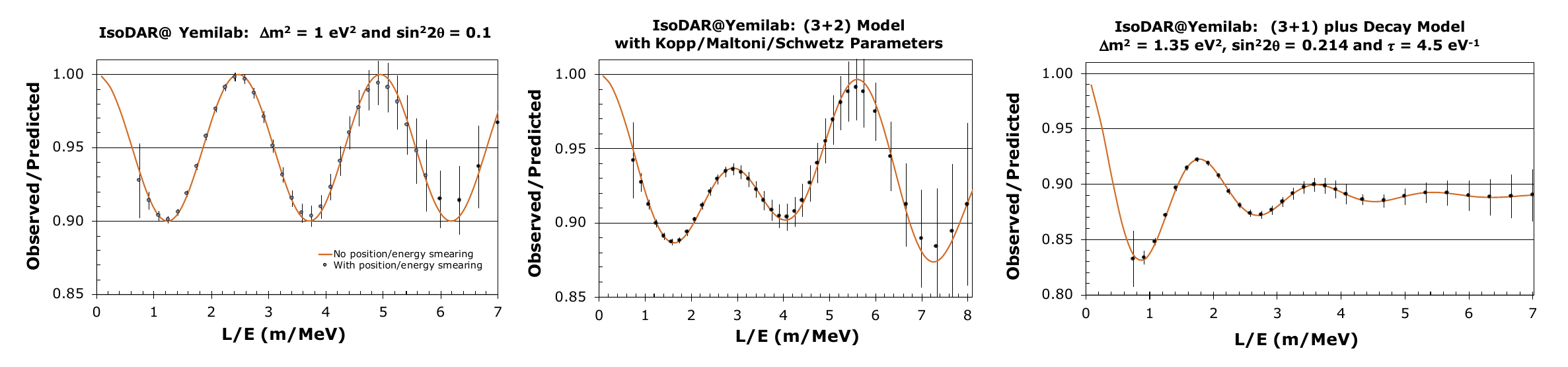}
\end{center}  
\vspace{-0.6cm}
\caption{\label{wiggles} The IsoDAR@Yemilab capability to measure oscillations under three example representative new physics scenarios: a 3+1 model (left), a 3+2 model (center), and a 3+1 with neutrino decay model consistent with the 95\% allowed region observed at IceCube (right)~\cite{Moulai:2021zey}. The points on the left and middle plots include position and energy smearing based on the expected Yemilab detector resolutions. The plot on the right does not include this smearing.}
\end{figure*}

As stated above, the IsoDAR@Yemilab configuration can expect $1.67\times 10^{6}$ detected IBD in 5~years of running, after accounting for detection efficiency, allowing IsoDAR@Yemilab to have unprecedented sensitivity to new physics models through the model-agnostic approach of analyzing the detected rate as a function of $L/E$.  
Fig.~\ref{wiggles} compares three scenarios that cannot be distinguished using existing global data.  The simplest model involving three active neutrinos is the 3+1 model, which produces a $\overline{\nu}_e$ oscillation wave as a function of $L/E$, with survival probability given by:
\begin{equation}
P_{\bar{\nu}_e \rightarrow \bar{\nu}_e} = 1-4(1-|U_{e4}|^2)|U_{e4}|^2\sin^2(1.27\Delta m^2_{41}L/E)~, \label{PUee}\\
\end{equation}
where $\Delta m^2_{41}$ is the mass-splitting between the fourth neutrino mass state and the three lighter neutrino states that are effectively degenerate, and $U_{e4}$ is the mixing matrix element that represents the electron flavor composition of the fourth mass state in the extended PMNS matrix.    Terms involving the latter are often simplified to an electron-flavor dependent mixing angle, such that the survival probability is given by:
\begin{equation}
P_{\bar{\nu}_e \rightarrow \bar{\nu}_e} = 1 - \sin^22\theta_{ee}\sin^2(1.27\Delta m^2_{41}L/E)~. \label{Pee}\\
\end{equation}
Motivated by the arbitrariness of assuming only one sterile neutrino and by tension between the observed experimental anomalies and limits, 3+2 models, with two sterile neutrinos, were introduced.  In this case, the survival probability is given by:
\begin{eqnarray}
P^{\rm 3+2}_{\bar{\nu}_e \rightarrow \bar{\nu}_e}&&= \nonumber \\
&& 1-4|U_{e4}|^2|U_{e5}|^2\sin^2(1.27\Delta m^2_{54}L/E)\nonumber \\
&& -4(1-|U_{e4}|^2-|U_{e5}|^2)(|U_{e4}|^2\sin^2(1.27\Delta m^2_{41}L/E) \nonumber \\
&& +|U_{e5}|^2\sin^2(1.27\Delta m^2_{51}L/E))~,  \label{Pee32}
\end{eqnarray}
where there is an additional mass splitting due to the fifth mass state, and the mixing matrix is further extended to include the coupling of the electron flavor to this state. 
Examples of the expected data as a function of $L/E$ for some characteristic 3+1 and  3+2~\cite{Kopp:2013vaa} IsoDAR@Yemilab scenarios are shown in Fig.~\ref{wiggles}, left and center.   One can see that, for IsoDAR@Yemilab, 3+2 is distinguishable from 3+1 due to the the interference between the two contributing mass splittings.      Fig.~\ref{wiggles} (right) presents the expectation for a representative ``3+1+decay'' scenario, a new model that has recently been  motivated by IceCube's muon-flavor disappearance results.   IceCube atmospheric muon neutrino data in the 1~TeV range will exhibit a resonant disappearance signature due to matter effects if neutrinos have a sterile component in the range of $\sim 1$~eV$^2$.   The results indicate an allowed region for a 3+1 fit at $>90\%$ and $<95\%$ CL~\cite{Moulai:2021zey}.    When the model is extended to allow for decay of the high mass neutrino $\nu_4$, the fit improves, and the SM is rejected with a p-value of 2.8\%~\cite{Moulai:2021zey}.   This motivates exploration of the model by IsoDAR, for the lifetime found by IceCube and $\Delta m^2$ within the IceCube 95\% allowed region that overlaps with a solution found in short-baseline global fits.   The survival probability for a 3+1+decay model is given by:
\begin{eqnarray}
P^{\rm 3+1+decay}_{\bar{\nu}_e \rightarrow \bar{\nu}_e}&&= \nonumber \\
&&    2 U_{e4}^{2} e^{-2.53 \frac{m  L}{\tau E}} (1 - U_{e4}^{2})\cos\left(2.53 \frac{L \Delta{m}^{2}}{E}\right) \nonumber \\
&& + U_{e4}^{4} e^{-5.07 \frac{m L}{\tau E}} + (1 - U_{e4}^{2})^{2}~.
\end{eqnarray}
Fig.~\ref{wiggles}, right, shows the IsoDAR rate as a function of $L/E$, and one can see the signature exponential die-off of the oscillation wave associated with the decay.

\begin{figure}[h]
\begin{centering}
\includegraphics[width=8.7cm]{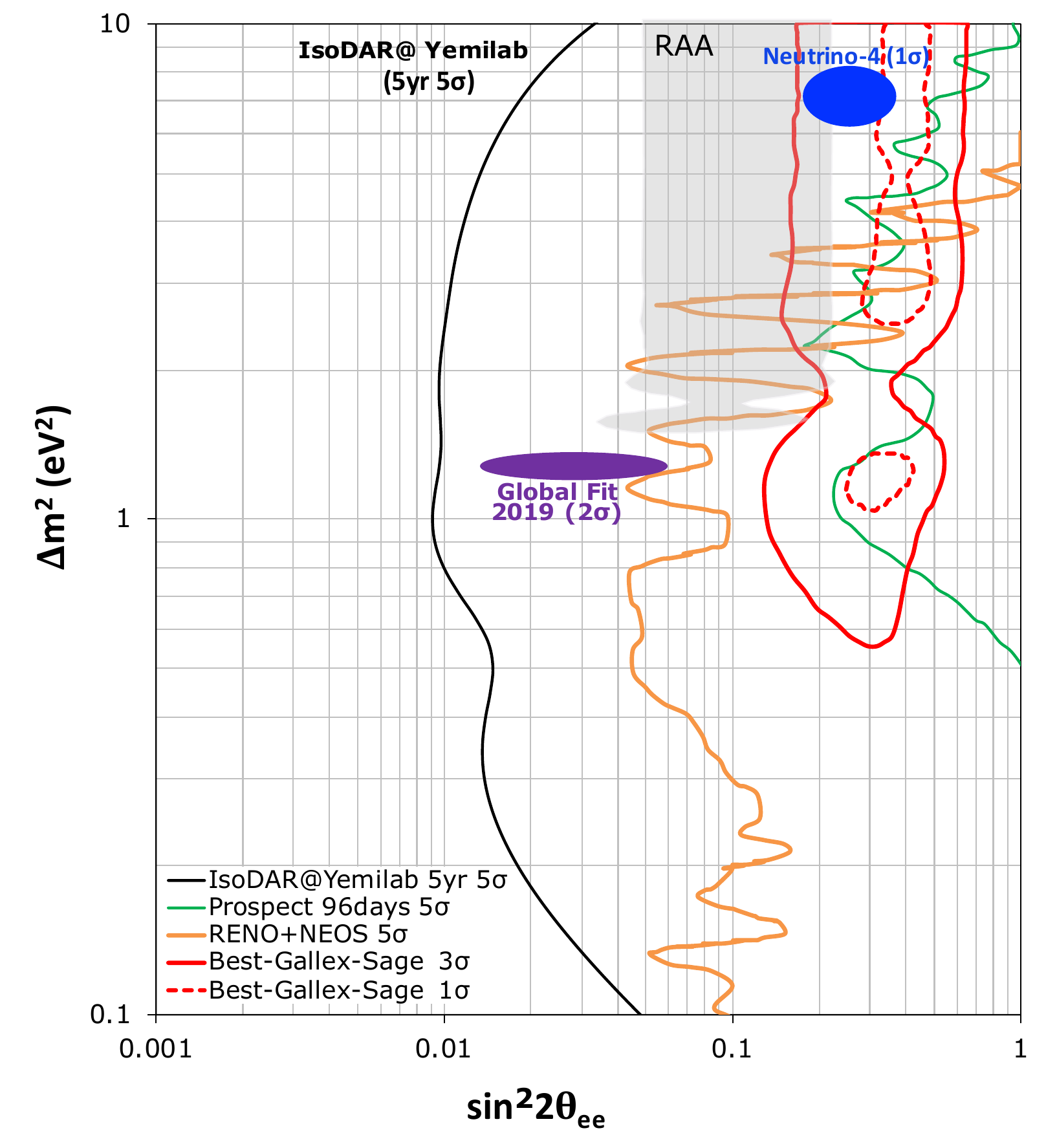}
\vspace{-.6cm}
\caption{The 5$\sigma$ sensitivity achievable by the IsoDAR@Yemilab experiment in 5 years of running, compared to a number of existing electron-flavor disappearance measurements.}
\label{sensitivity}
\end{centering}
\end{figure}

The sensitivity to $\bar{\nu}_e \rightarrow \bar{\nu}_{e}$ is traditionally calculated and compared with existing data within a 3+1 model, using Eq.~\ref{Pee}. The specifics of the IsoDAR sensitivity calculation, based on searching for $L/E$ shape-dependent effects, follows Ref.~\cite{Agarwalla:2011qf}. Fig.~\ref{sensitivity} shows the IsoDAR@Yemilab 5$\sigma$ sensitivity for 5 years of running, as described above.

Fig.~\ref{sensitivity} also demonstrates the present state of electron-flavor disappearance searches, which is very complex.   One sees four closed contours, which are allowed regions:  the Reactor Antineutrino Anomaly (gray)~\cite{Mention:2011rk},  the Neutrino-4 reactor experiment measurement (blue)~\cite{Serebrov:2020kmd}, a BEST-GALLEX-SAGE~\cite{Barinov:2021asz,Barinov:2022wfh,Kaether:2010ag,SAGE:2009eeu} source experiment combination (red)~\cite{Serebrov:2020kmd}, and a 2019 global fit (purple)~\cite{Diaz:2019fwt}.  One immediately notes that the allowed regions have significant disagreements.  Also, a set of recent reactor experiments have not observed $\bar \nu_e$ disappearance, and therefore set corresponding limits;  examples on Fig.~\ref{sensitivity} are PROSPECT (green)~\cite{PROSPECT:2020sxr} and the combined NEOS/RENO analysis (yellow)~\cite{RENO:2020hva}. The 5$\sigma$ exclusion curve for NEOS/RENO (PROSPECT) is obtained by extracting the $\sin^2\theta_{41}$ and $\Delta m^2_{41}$ values from the 90\% (95\%) CL exclusion curve in Ref.~\cite{RENO:2020hva} (Ref.~\cite{PROSPECT:2020sxr}) and then multiplying $3.05=5\sigma/1.64\sigma$, ($2.55=5\sigma/1.96\sigma$) to the $\sin^2\theta_{14}$ values. As can be seen, these limits are in strong disagreement with the low-$\Delta m^2$ solution for BEST-GALLEX-SAGE.   

The lack of clarity in Fig.~\ref{sensitivity} indicates that a simple 3+1 oscillation model is unlikely to explain all of these results.    Either one or more of the results is incorrect or the underlying physics is significantly more complicated than a 3+1 model.   
However, as can be seen in Fig.~\ref{sensitivity}, the IsoDAR@Yemilab experiment will cover these results at very high sensitivity. In general, if \textit{any} or all signatures are real, they will be easily discernible with IsoDAR@Yemilab.   Beyond this, the IsoDAR@Yemilab design has several features that make it ideal to follow up on these experiments: 
\begin{itemize}
    \item Unlike reactor experiments,  the IsoDAR flux is created by a \textit{single} isotope ($^8$Li beta decay), which is extremely well understood.
    \item BEST-GALLEX-SAGE are MegaCurie single-isotope source experiments. However, these experiments only count germanium atoms produced from $\nu_e$ charged current interactions, and cannot reconstruct individual events.   IsoDAR$@$Yemilab reconstructs the neutrino path length, $L$, and energy, $E$, on a \textit{per-event} basis.
    \item The IsoDAR $L$, $E$, and $L/E$ ranges are uniquely wide and the $L$ and $E$ reconstruction capability is at high precision compared to the other experiments.  
\end{itemize}
It is worth noting that most of the anomalies discussed are the result of follow-up on previous experiments of the same type. Unfortunately, if history is any kind of predictor of the future, running more of the same type of experiment, with incremental (and, even substantial) improvements, is unlikely to provide a definitive explanation of the complicated and confusing situation. IsoDAR@Yemilab provides a new way to explore the problem with both unprecedented sensitivity and approach. Indeed, in the case that any one of the existing anomalies is due to some kind of new physics involving oscillations, IsoDAR will almost certainly make a discovery.

\section{Oscillation measurements in a global context \label{global}}

If the possible source of new physics is due to the existence of one or more sterile neutrinos, then the IsoDAR$@$Yemilab $\bar \nu_e$ disappearance sensitivity requires consideration within the context of global searches for $\nu_\mu$ disappearance and $\nu_\mu \rightarrow \nu_e$ appearance (noting that oscillations involving $\nu_\tau$ are weakly constrained at present).  This is because the addition of extra mass and flavor states into the neutrino sector leads to all three phenomena.  Returning to a 3+1 model for simplicity, while keeping in mind that the tension between various experiments indicates that this model is likely to be too simplistic (if sterile neutrinos do exist), then one finds two more equations that make a triplet with Eq.~\ref{PUee}:
\begin{eqnarray}
P_{\nu_\mu \rightarrow \nu_\mu}& \cong & 1-4(1-|U_{\mu 4}|^2)|U_{\mu 4}|^2\sin^2\left(\frac{1.27\Delta m^2_{41}L}{E}\right)\label{PUmu4} \nonumber\\
P_{\nu_{\mu}\rightarrow\nu_e}& \cong &  4|U_{e4}|^2 |U_{\mu4}|^2\sin^2\left(\frac{1.27\Delta m^2_{41}L}{E}\right)\label{PUe4mu4}
\end{eqnarray}
where, for clarity, we write out the extended PMNS matrix for 3+1:
\begin{equation}
U_{3+1} = \begin{bmatrix}
U_{e1} & U_{e2} & U_{e3} & U_{e4} \\ 
\vdots & & \vdots & U_{\mu4} \\
\vdots & & \vdots & U_{\tau4} \\
U_{s1} & U_{s2} & U_{s3} & U_{s4}
\end{bmatrix} \label{4mixmx}
\end{equation}
By making the mixing matrix elements explicit, one sees the connection between disappearance data sets ($P_{\nu_e \rightarrow \nu_e}$ and $P_{\nu_\mu \rightarrow \nu_\mu}$) and appearance data sets ($P_{\nu_{\mu}\rightarrow\nu_e}$) in a 3+1 model.

Since 1993, possible $3+N$ ($N>0$) oscillation effects have been observed in all three channels, although the $\nu_\mu$ hints are very recent~\cite{IceCube} and do not rise to the $>2\sigma$ level.  At the same time, the parameter space is highly constrained in global, all-experiment fits by results that have reported no oscillation signature, thus the allowed region for the global fit (purple) on Fig.~\ref{sensitivity} is rather small.
Notably, there is a well-known ``tension'' between the disappearance and appearance data sets, which is traditionally quantified using the ``parameter goodness-of-fit" (PG) test~\cite{Maltoni:2003cu}.   By that measure, there is a ${3.7\cdot 10^{-6}}$ probability of agreement between the worldwide short-baseline data sets within a 3+1 model~\cite{Diaz:2019fwt}.   On the other hand, the $\Delta \chi^2$/$dof$ when fitting the data for a 3+1 model versus an only-3 model shows a $5.2\sigma$ improvement with the addition of the sterile state~\cite{Diaz:2019fwt}. This indicates that, although 3+1 may not be the correct underlying model, the data strongly prefer a model with oscillations involving a sterile neutrino over the SM.   



The MiniBooNE experiment, which uses a Cherenkov detector, has observed an excess of $\nu_e$-like events for $E_\nu<500$ MeV at a baseline of $\sim$550~m.   To date, most 3+1  studies involving MiniBooNE have assumed that $\nu_\mu \rightarrow \nu_e$ appearance may occur, but do not also consider that $\nu_e$, which is also a component of the intrinsic decay-in-flight flux, can also disappear.  Very recently, Kopp and Brdar~\cite{Brdar:2021cgb} have studied the allowed regions for MiniBooNE alone, outside of a global fit, allowing all three oscillation modes associated with sterile neutrinos.    This is a complicated fit, and an external, very high statistics measurement of oscillation parameters from IsoDAR@Yemilab would allow the intrinsic $\nu_e$ prediction at MiniBooNE to be well determined, thereby allowing the appearance parameters to be extracted.   Given a precise oscillation prediction, we can isolate any remaining excess MiniBooNE signal, which could be due to photons.  This is inconsistent with a simple 3+1 model that says 100\% of the MiniBooNE signal is due to $\nu_e$ interactions.

The need for IsoDAR@Yemilab has become even more apparent because of the recent MicroBooNE results~\cite{MicroBooNE:2021rmx}.  MicroBooNE is located 70~m upstream of MiniBooNE and uses liquid argon time projection chamber (LArTPC) technology.   This state-of-the-art detector can distinguish electrons from photons, unlike the MiniBooNE detector.  MicroBooNE limits the fraction of generic $\nu_e$ charged-current interactions that can explain the MiniBooNE excess to $<51\%$ at 95\% CL.  However, the limit is significantly better than the expected experimental sensitivity because MicroBooNE observes an overall deficit of $\nu_e$, which may be a complex disappearance signature, perhaps similar to those that produce the $L/E$ dependence shown in Fig.~\ref{wiggles}. This possibility is already being explored in the literature (see, e.g., Ref.~\cite{Denton:2021czb}). Alternatively, because MicroBooNE is using state-of-the-art detection technology, it may be that there are unmodelled inefficiencies which lead to the deficit.  

The combination of the apparent cross-experiment disagreement with electron-flavor disappearance, the tension between appearance and disappearance in the global fits, and the possible unmodeled source of photon-like signals in MiniBooNE, point to a more complex sterile neutrino model, if new physics is indeed the source of the anomalous results.
In Fig.~\ref{wiggles} (middle), we show the IsoDAR@Yemilab sensitivity to an additional sterile neutrino state (a ``3+2'' model).
Notably, the addition of the second new state allows for new sources of $CP$ violation to be present in appearance results. Therefore, the ``nuisance" of sterile-neutrino-induced $CP$ violation is required in global fits for a mixed conventional beam that has neutrinos and antineutrinos. On the other hand, the fact that we are performing a disappearance search using a pure electron-antineutrino source implies that IsoDAR@Yemilab is insensitive and agnostic to the value of the $CP$ phases. Unfortunately, at present, a $3+2$ model does not seem to significantly reduce the tension between appearance and disappearance experiments in global fits, and there is no compelling improvement overall, as compared to a $3+1$ model~\cite{Diaz:2019fwt}.

A possible solution to the tension is to introduce additional secret forces that predominantly affect the mostly sterile neutrino mass state. Indeed, this might produce an additional photon-like signal in MiniBooNE.
In this context, the possibility of the mostly sterile neutrino mass state to decay has been considered in Refs.~\cite{decay1,decay2,decay3,decay4,decay5,decay6,decay7,decay8,decay9,decay10,decay11}. Notably, in Ref.~\cite{Moulai:2019gpi} it was shown that a 3+1+decay model significantly reduces the tension between appearance and disappearance experiments, improving the global-data goodness-of-fit. Sterile neutrino decay leads to a dampening in the neutrino oscillation pattern as can be seen in Fig.~\ref{wiggles}, right, which produces a clear and distinct signature. In general, IsoDAR has excellent sensitivity to the IceCube-motivated parameters of a 3+1+decay model. 

\section{Sensitivity to wavepacket effects \label{nuewave}}

Adding to the complexity of oscillation searches at short-baseline, the authors of Ref.~\cite{ArguellesSalvado} have recently pointed out that the assumption of the neutrino state as a plane wave (PW) may be too simplistic for oscillation models applied to the $\Delta m^2\sim1~\mathrm{eV}^2$ anomalies.  A better description is the wave package (WP) formalism, which accounts for effects arising from a finite initial antineutrino wavepacket width.  This leads to decoherence that becomes most apparent when $L/E$ is large compared to $1/\Delta m^2$.  The wavepacket width depends on the source, and for reactors could be finite due to the characteristic sizes of the quantum system, including U and Pu nuclei ($10^{-5}$~nm), the inverse of the antineutrino energy ($10^{-4}$~nm) and the interatomic spacing ($10^{-1}$~nm)~\cite{deGouvea:2021uvg}.  Fits to a combination of Daya Bay, RENO and KamLAND data set a limit on the wave-packet width of $\sigma_x > 2.1 \times 10^{-4}$ nm at 90\% CL~\cite{deGouvea:2021uvg}.  

\begin{figure}[t]
\begin{centering}
\includegraphics[width=8.3cm]{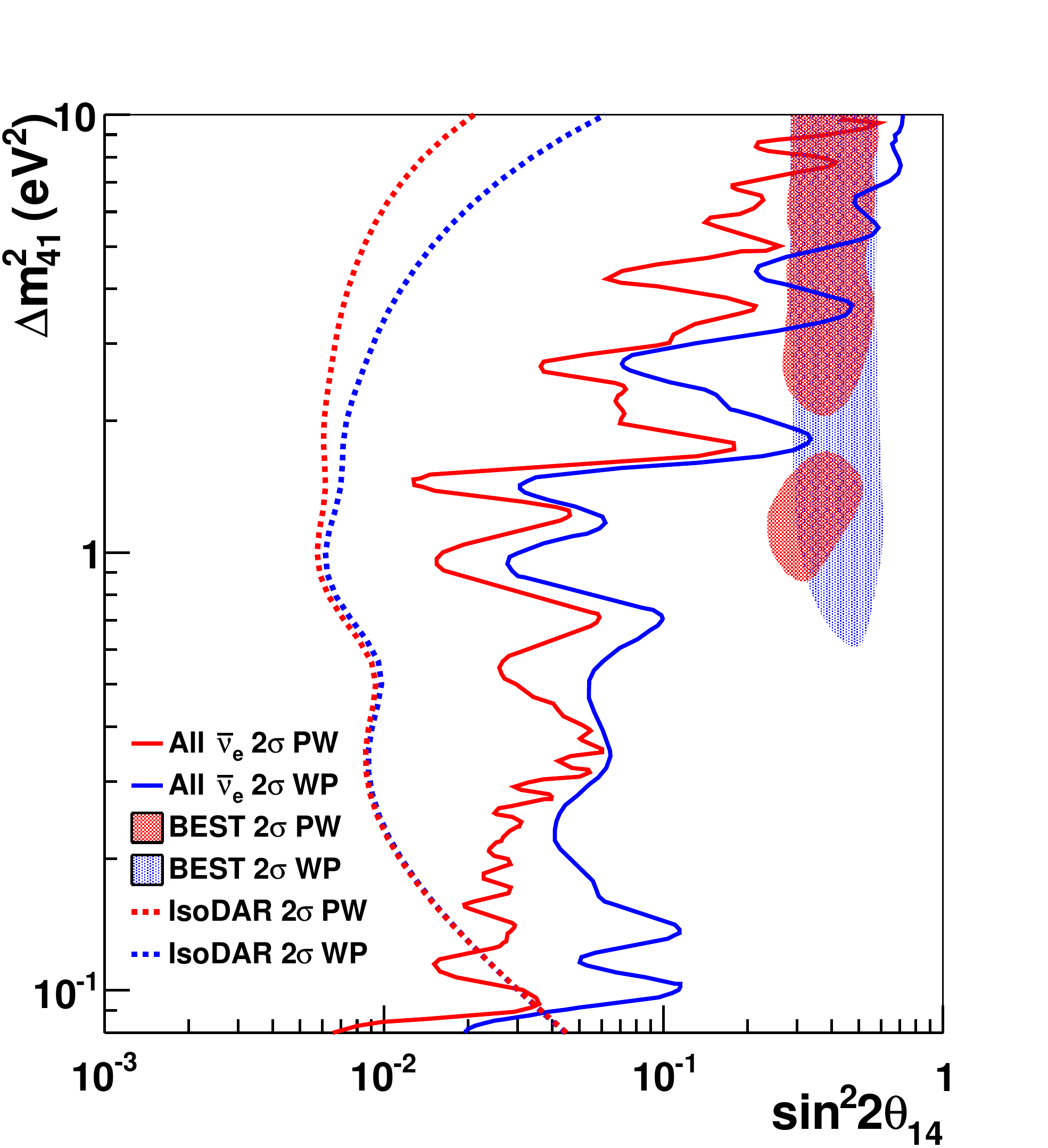}
\vspace{-.1cm}
\caption{The 2$\sigma$ IsoDAR (4~year livetime) sensitivity for a wavepacket (WP) effect  of $\sigma_x = 2.1 \times 10^{-4}$~nm as compared to the nominal plane-wave (PW) sensivity. The results from BEST alone and a combination of Daya-Bay-NEOS-PROSPECT (``All $\overline{\nu}_e$") are also shown. This figure has been adapted from Ref.~\cite{ArguellesSalvado}.} 
\label{best_plot}
\end{centering}
\end{figure}

The wavepacket width is incorporated into the 3+1 electron-flavor survival probability equation in the following manner~\cite{ArguellesSalvado}:
\begin{equation}
    P_{ee}^{WP} = 1 - \sin^2 2\theta \Big[ \big( {{1-e^{-A^2}}\over{2}} \big)+\sin^2 \big({{1.27 \Delta m^2 L}\over{E}}\big) e^{-A^2} \Big]~,
\end{equation}
where $A= L/L^{coh}$ and 
\begin{equation}
    L^{coh}=5.627 \times 10^{12} \big( {{E^2 \sigma_x}\over{\Delta m^2}} \big)~.
\end{equation}
In the above equations, $L$, $L^{coh}$ and $\sigma_x$ are in meters, and $E$ is in MeV.    Fits assuming $\sigma_x = 2.1 \times 10^{-4}$~nm lead to considerably reduced limits in the high-$\Delta m^2$ anomaly range for a combined fit to Daya Bay, RENO, and PROSPECT~\cite{ArguellesSalvado}. The MegaCurie $^{51}$Cr and $^{37}$Ar source experiments, BEST, GALLEX and SAGE, would be expected to have a wavepacket effect of roughly similar level.   If one fits the source data using $\sigma_x = 2.1 \times 10^{-4}$~nm, the allowed region is enlarged in the $\sim 1.5-2$ eV$^2$ region, as can be seen in Fig.~\ref{best_plot}. The figure also shows that a wavepacket effect of $\sigma_x = 2.1 \times 10^{-4}$~nm leads to $>2\sigma$ agreement between reactors (``All $\overline{\nu}_e$" in the figure) and BEST at $\Delta m^2 \sim1.7$~eV$^2$ as well as $>5$~eV$^2$, hence, dramatically improving tension between the experiments.

\begin{figure}[t]
\begin{centering}
\includegraphics[width=8.8cm]{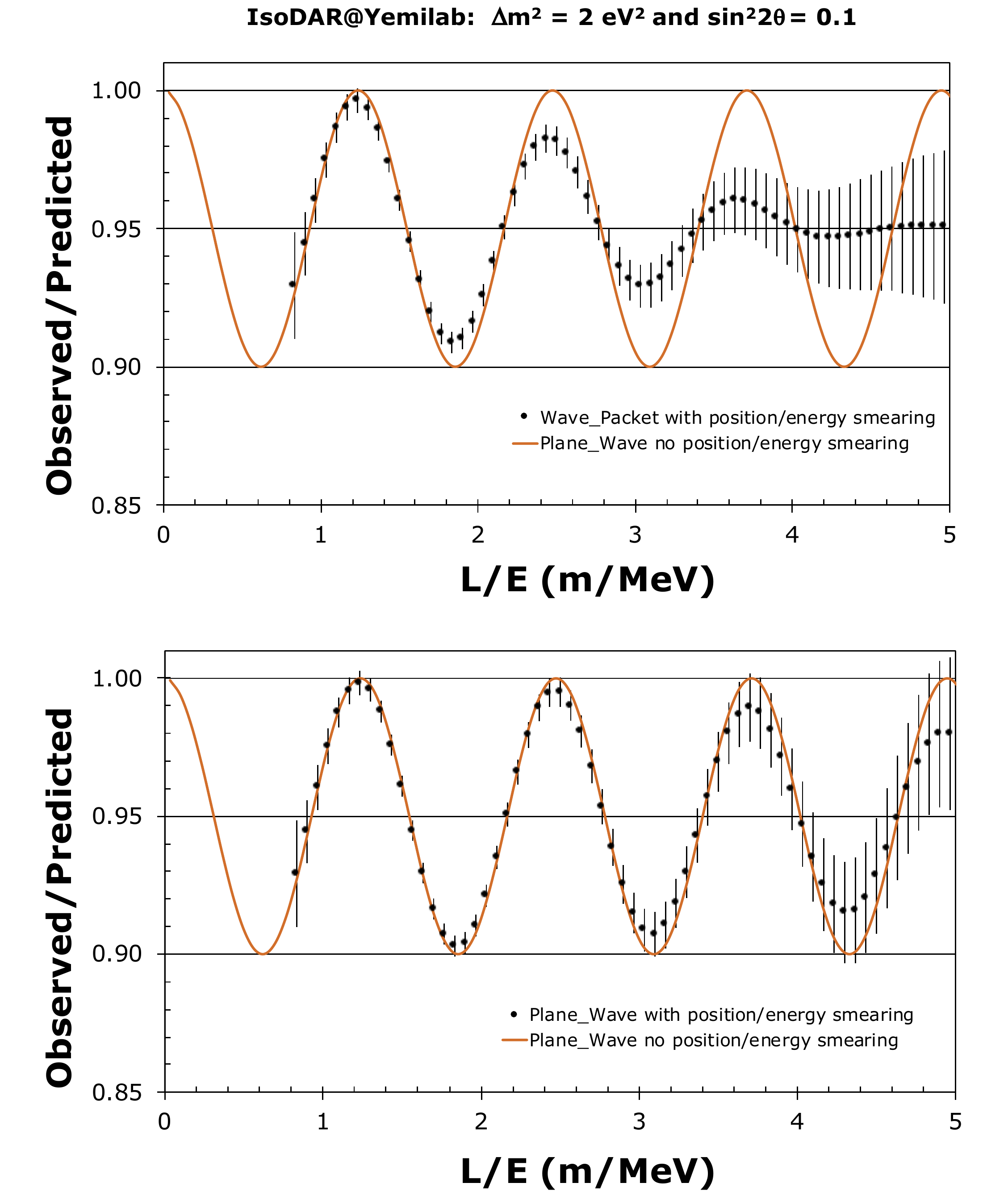}
\vspace{-.3cm}
\caption{Top: IsoDAR@Yemilab 4~year livetime rate versus $L/E$ for a 3+1 example with a wavepacket effect of $\sigma_x = 2.1 \times 10^{-4}$ nm and $\Delta m^2=2~\mathrm{eV}^2$;  Bottom:  the same, but with no wavepacket effect.} 
\label{wigglesWP}
\end{centering}
\end{figure}

The high statistics of the IBD data set allows IsoDAR@Yemilab to study the wavepacket effect.  Since the bulk of this data will be in the range where $L/E$ (in m/MeV) is not substantially larger than the $\Delta m^2$ (in eV$^2$) range under scrutiny, the sensitivity is not significantly affected.  Fig.~\ref{best_plot} shows the 2$\sigma$ IsoDAR@Yemilab sensitivity for a wavepacket effect  of $\sigma_x = 2.1 \times 10^{-4}$~nm, alongside the nominal plane-wave sensitivity and, adapted from Ref.~\cite{ArguellesSalvado}, the analogous results from both BEST individually and a Daya-Bay-NEOS-PROSPECT experiment combination~\cite{DayaBay:2016qvc,DayaBay:2016ggj,NEOS:2016wee,PROSPECT:2020sxr}. For IsoDAR@Yemilab, one can see that at $\Delta m^2=1$ eV$^2$ the difference in sensitivity is $\sim$10\%.  This sensitivity can be restored through a minor extension to the run-time. However, at the same time, the experiment has sufficient statistical power at large $L/E$, where the effect is largest, that the data can be fit for the value of $\sigma_x$.  Fig.~\ref{wigglesWP} shows the $L/E$-dependence of the rates for $\Delta m^2=2$~eV$^2$ oscillations with $\sigma_x = 2.1 \times 10^{-4}$~nm (top) and with no wavepacket effect (bottom).  For $L/E>3$~m/MeV, the wavepacket effect is readily apparent with the data distribution substantially flattened.  The ability to distinguish these oscillation models and measure the wavepacket effect at high precision is unique to IsoDAR@Yemilab.

\section{Searches for new physics via ``bump hunting'' in the IBD spectrum}

\begin{figure}[h]
\begin{centering}
\includegraphics[width=8.7cm]{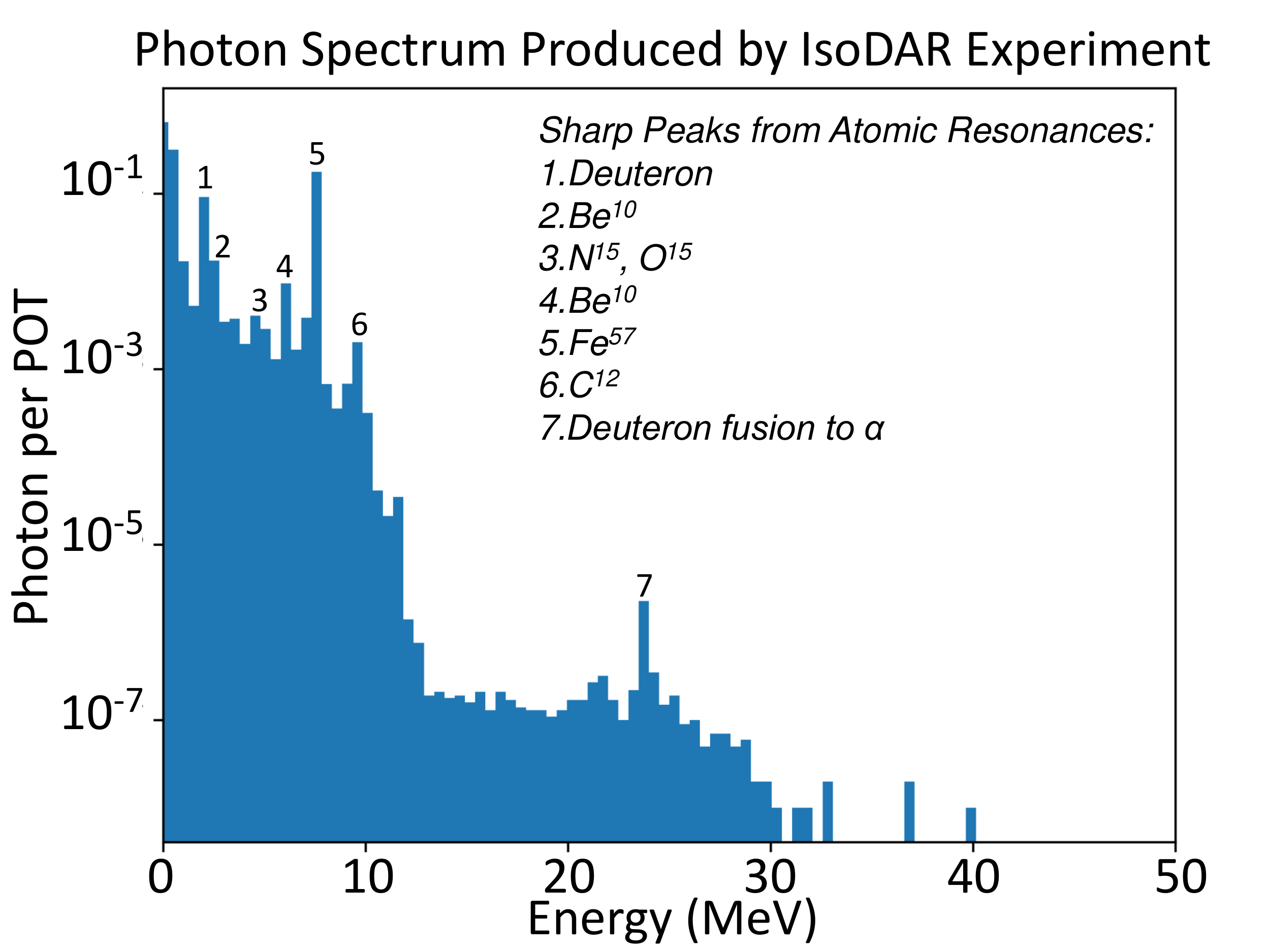}
\vspace{-.6cm}
\caption{The photon spectrum expected from the IsoDAR target modeled using the \texttt{QGSP\_BIC\_ALLHP} library in Geant4~\cite{GEANT4:2002zbu}. This spectrum is used to calculate the achievable $X$-decay sensitivity shown in Fig.~\ref{x_nunu_sensitivity}.}
\label{photons}
\end{centering}
\end{figure}

The same excellent energy resolution, seen in Fig.~\ref{ratesforIBD}, that allows precise searches for deficits related to oscillations and/or decays involving sterile neutrinos also allows for searches for peaks from new particles produced in the IsoDAR target and sleeve that decay to $\nu_e \bar \nu_e$.   There is both theoretical and experimental impetus for a ``bump hunt'' at IsoDAR@Yemilab.  The theoretical motivation arises from interest in low mass mediators, called \emph{light X particles}
, that are produced through mixing of photons within the target and sleeve or directly from the nuclear transitions emerging at the target or the sleeve.  Fig.~\ref{photons} shows the spectrum of photons that are produced up to 50 MeV, where the line-structure arises from the transitions of excited nuclei, and the other photons are mainly due to bremsstrahlung. If the $X$ 
particles are nearly at rest, the subsequent decays produce $\bar \nu$s with energy at half the mass, which can engage in an IBD interaction, producing a peak.     The experimental motivations are twofold: the Atomki anomaly and the 5 MeV ``reactor bump.''  We discuss all motivations below, however we note that, given the novel high-statistics data set that will be produced by IsoDAR@Yemilab, a curiosity-driven search is equally valid.

Low mass mediators are well motivated in various extensions of the SM, see e.g., Refs.~\cite{Bauer:2018onh, Dutta:2019fxn, Datta:2018xty,DelleRose:2018eic,Dutta:2021cip}. These models involve extensions of the SM gauge sectors and/or the SM Higgs sector and are motivated to explain the origin of dark matter, neutrino masses and mixings, non-standard neutrino interactions and various anomalies, e.g., $g-2$ of the muon, Atomki, MiniBooNE, LHCb, etc. The mediators can be of vector, scalar and pseudoscalar types and involve  couplings to the SM and dark sector particles. Various beam dump experiments, Belle, BaBar, reactor and beam-dump based neutrino experiments, astrophysical measurements, etc. apply constraints on these mediators. The  mediators of mass $O(10)$~MeV can  provide positive and negative contributions to $\Delta N_{eff}$ depending on its decay branching ratio into neutrino-anti-neutrinos and electron-positrons~\cite{Escudero:2018mvt}, respectively, which is interesting to determine the allowed parameter space for these low mass mediator models. Further, the interactions involving the decay into neutrino final states have impact on the neutrino floor for dark matter direct detection experiments~\cite{AristizabalSierra:2019ykk, Boehm:2018sux}.  


The transition lines in Fig.~\ref{photons} are associated with various types of  magnetic ($M_i$) and electric ($E_i$) moments which can be associated with different types of mediators, e.g.,~\cite{Avignone:1988bv,Feng:2016ysn,Dent:2021jnf}. Due to the existence of many lines with different moments, for simplicity, we assume that the generic mediator  $X$ is coupled to both quarks and neutrinos, e.g., ~\cite{Datta:2018xty,AristizabalSierra:2019ykk, Boehm:2018sux}.  The production rate of this new mediator depends on its coupling with quarks and the mass, which can be expressed as a branching ratio for a given transition. This simple model  can bypass the constraints from the electron beam dump data, but  the product of the neutrino and quark couplings is limited by some neutrino experiments, e.g., COHERENT, CCM etc., and has impacts on the neutrino floor for direct detection experiments~\cite{Boehm:2018sux, AristizabalSierra:2019ykk}.

\begin{figure}[h]
\begin{centering}
\includegraphics[width=8.7cm]{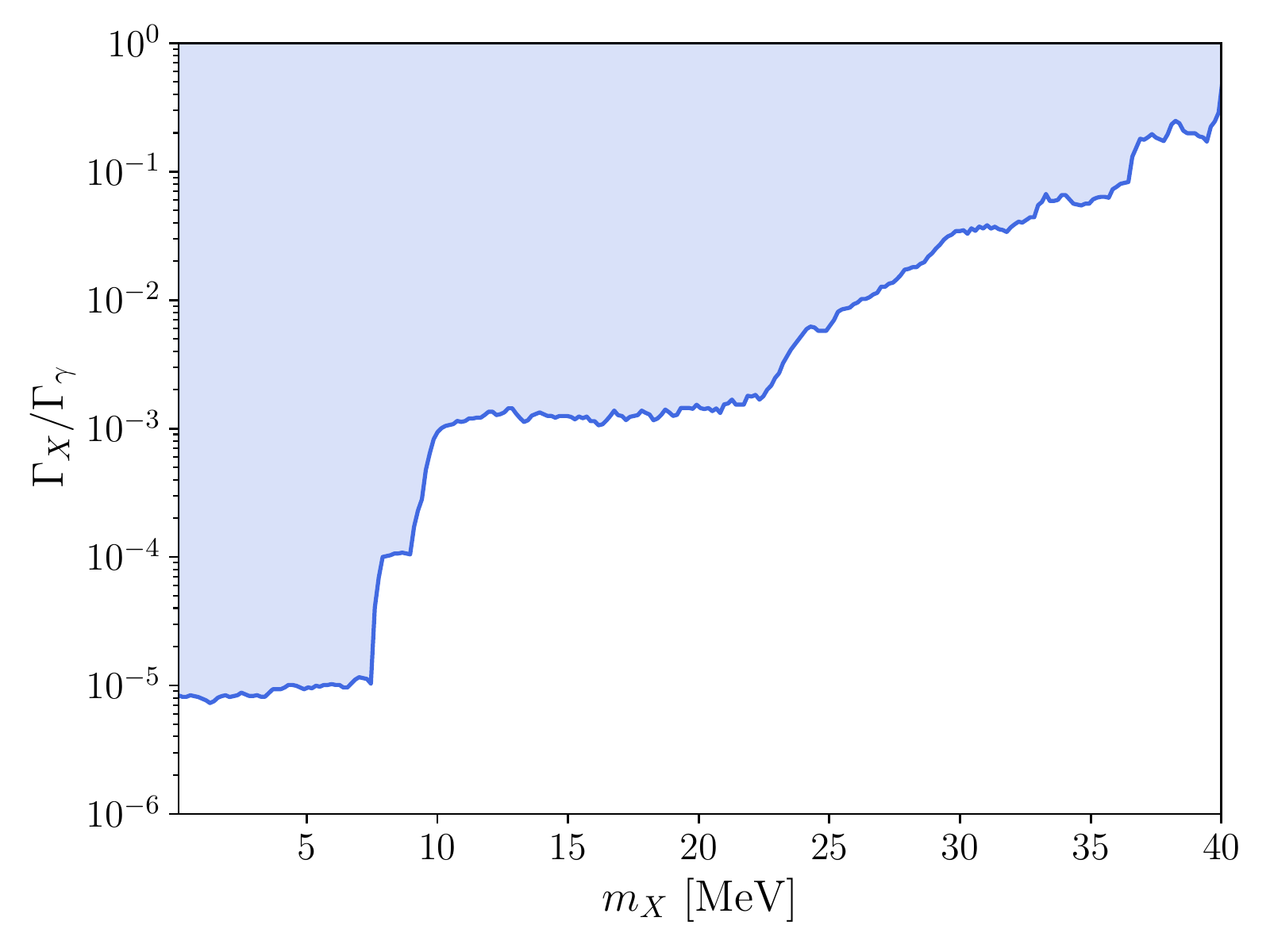}
\vspace{-.6cm}
\caption{The IsoDAR 4-year livetime sensitivity exclusion on the $N^* \to N X(\to \bar{\nu}\nu)$ branching ratio as a function of the boson mass $m_X$, given at 90\% CL. The flat limit for $m_X \lesssim 5$ MeV may extend to arbitrarily small masses (sub-eV) barring model-dependent bounds.}
\label{x_nunu_sensitivity}
\end{centering}
\end{figure}

\begin{figure}[h]
\begin{centering}
\includegraphics[width=8.8cm]{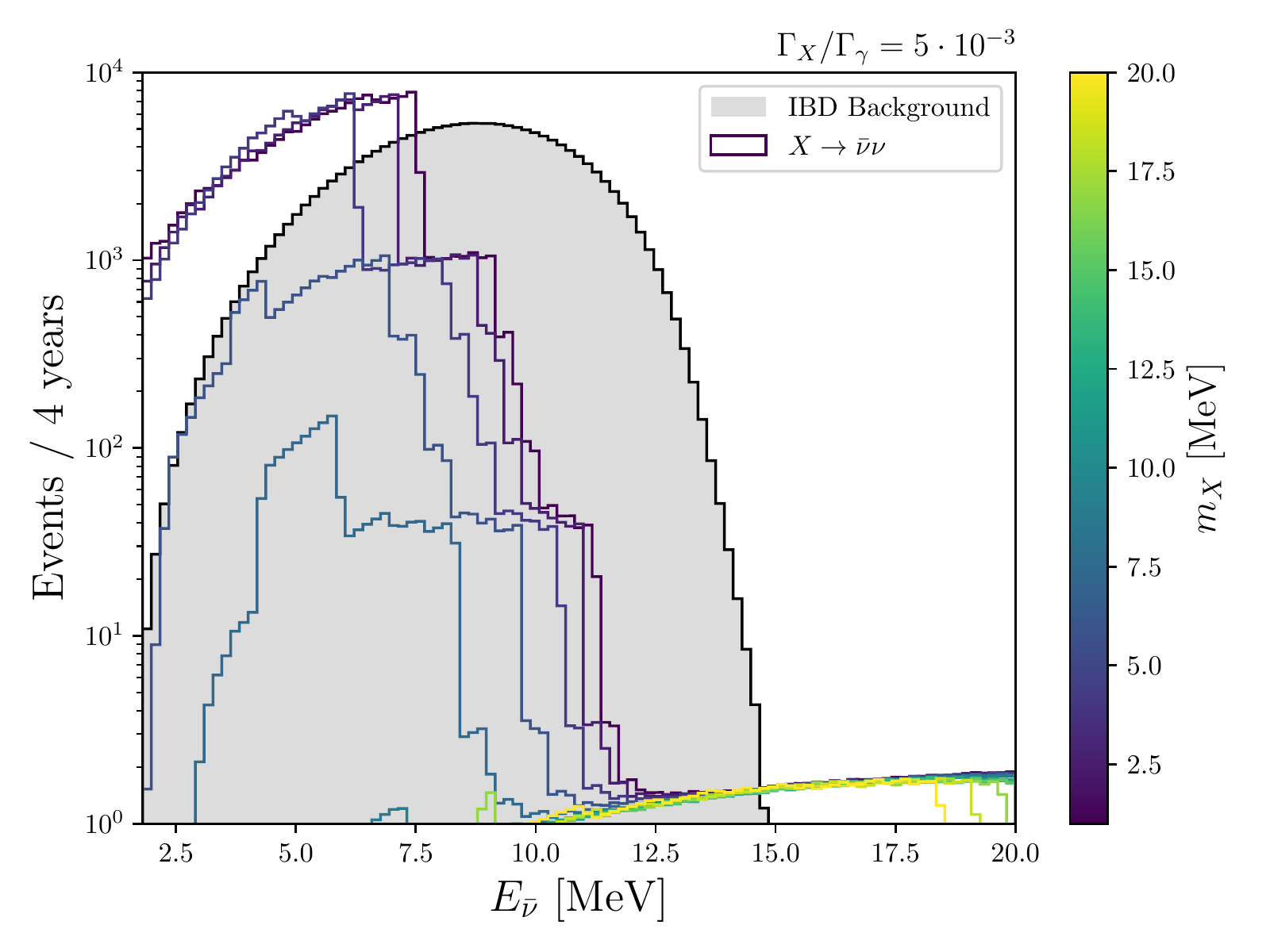}
\vspace{-.6cm}
\caption{The IBD ($\bar{\nu}_e + p \to e^+ + n$) rates from promptly decaying $X\to\bar{\nu}_e\nu_e$ arriving at the Yemilab detector (color-coded by mass) plotted against the expected IBD background (gray). The spectral shape is inherited from the convolution of the boosted 2-body decay spectrum with the IBD cross section, $\sigma_\textrm{IBD}(E_{\bar{\nu}_e})$, and summed over all kinematically accessible nuclear transitions to produce the $X$ states.}
\label{x_nunu_spectra}
\end{centering}
\end{figure}


Driven by nuclear transition induced gammas in the IsoDAR target, we present the 90\%~CL sensitivity to the bosonic state $X$ via a branching ratio $\Gamma_X / \Gamma_\gamma$, that subsequently decays to $\bar{\nu}\nu$ pairs, in Fig.~\ref{x_nunu_sensitivity}. The $\overline{\nu}_e$ spectrum produced from the prompt $X$ decay is simulated with Monte Carlo in the $X$ rest frame, boosted to the lab frame, and propagated to the IsoDAR@Yemilab 2.26~kton fiducial volume where the antineutrino is detected via IBD. The $\bar{\nu}_e$ energy spectra detected this way are shown in Fig.~\ref{x_nunu_spectra} for several masses and compared with the IBD rate from $^{8}$Li, which is expected to be the only significant background for this search. There are several interesting features of the signal shape, namely, the boosted $\bar{\nu}_e$ spectrum from each monoenergetic $X$ produced would have endpoint energies $E_{\bar{\nu}_e}^{\substack{max\\min}} = \gamma m_X(1 \pm \beta)/2$, where $\gamma$ and $\beta$ are the Lorentz factor and $X$ velocity, respectively. The edges in the spectrum are the imprints of edges in the photon spectrum (Fig.~\ref{photons}), transformed and skewed by the combined effect of the Lorentz boost and IBD cross section convolution.

Our projected sensitivity in Fig.~\ref{x_nunu_sensitivity} is then calculated by performing a $\Delta\chi^2$ analysis, treating the expected IBD spectrum as a background and null hypothesis. The $X$-boson coupling to the quarks can be constrained from this analysis to be $\leq 10^{-3}$ for an $X$-boson mass $O(10)$~MeV 
when the $X$ boson decays promptly into neutrinos with coupling values $\geq 10^{-7}$. Some regions of the parameter space associated with the product of the quark and the neutrino couplings of $X$ that can be probed at IsoDAR are still allowed by the   constraints from the COHERENT experiment~\cite{AristizabalSierra:2019ykk,Cadeddu:2020nbr,us}. 

One particular experimental interest is the sensitivity to the light mediator claimed to explain the Atomki anomaly~\cite{Krasznahorkay:2015iga, Gulyas:2015mia, Aleksejevs:2021zjw, DelleRose:2018pgm}. This is a reported excess of $e^+e^-$ pairs observed in the decay of the 18 MeV excited state of beryllium produced through $^7$Li(p,n)$^8$Be$^*$, and the set of 20 MeV excited states of helium produced through $^3$H(p,$\gamma$)$^4$He.
In the former case, the invariant mass of the pairs is consistent with a vector boson mediator of 
$16.70\pm0.35$(stat)$\pm 0.5$(sys)~MeV and in the latter of mass $16.94\pm0.12$(stat)$\pm 0.21$(sys)~MeV. 
However, one can see from Fig.~\ref{photons}, that the rate of 18~MeV (and higher energy) photon production is relatively low. Thus, if IsoDAR observes a peak due to $\bar \nu_e$ interactions at $\sim$8.5~MeV, then the connection to the Atomki anomaly requires a coupling to neutrinos that is substantially different from the coupling to electrons. Alternatively, if IsoDAR@Yemilab observes no signal at 8.5~MeV, some (but unlikely all) explanations for the Atomki anomaly can be excluded.

\begin{figure}[h]
\begin{centering}
\includegraphics[width=8.6cm]{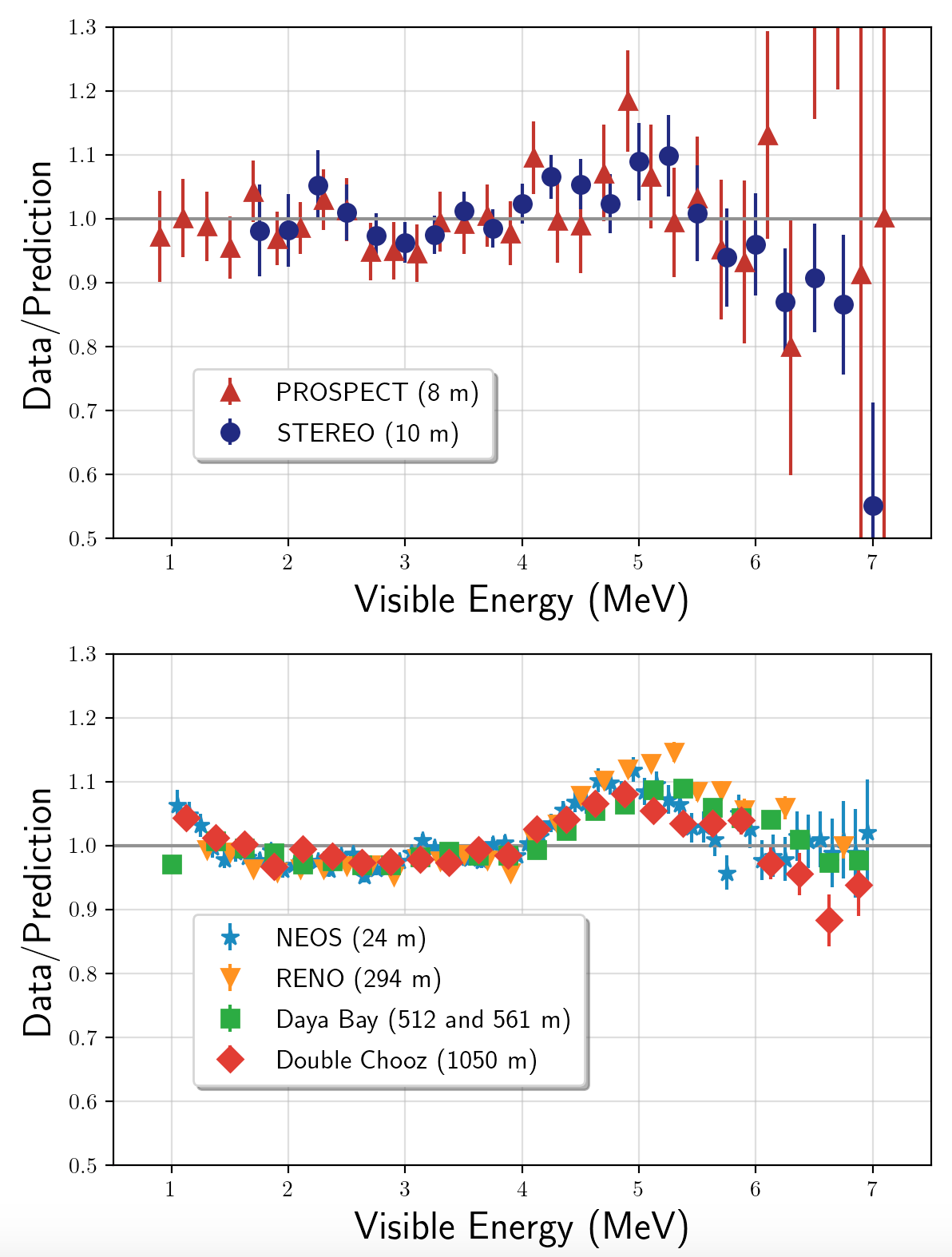}
\vspace{-.6cm}
\caption{The ratio of data to predicted IBD rate for reactor experiments versus visible energy, showing the 5 MeV excess.     Top: for detectors located at HEU reactors;   Bottom: for detectors located at power reactors. These plots have been updated from Ref.~\cite{Diaz:2019fwt}.}
\label{reactorbumps}
\end{centering}
\end{figure}

Another interesting experimental motivation arises from the 5~MeV reactor bump, which is seen in the event distribution of most modern reactor experiments. Fig.~\ref{reactorbumps} shows the ratio of data to prediction for recent high statistics data sets, with experiments located at highly enriched uranium (HEU) reactors, PROSPECT~\cite{prospect2} and STEREO~\cite{stereo}, shown in the top panel and those located at power-reactors, NEOS~\cite{NEOS:2016wee}, RENO~\cite{reno}, Daya Bay~\cite{dayabay} and Double Chooz~\cite{doublechooz}, in the bottom panel. The source of the excess at 5~MeV has not yet been fully explained, although recent measurements~\cite{RENO:2018pwo, DayaBay:2019yxq} indicate that the bump may arise from incorrect predictions of the Huber-Mueller model~\cite{Huber:2011wv,Mueller:2011nm} related to $^{235}$U (and perhaps $^{239}$Pu as well) isotopes in reactor cores. If that is the case, since IsoDAR uses an $^8$Li-decay flux, IsoDAR@Yemilab will not observe a 5~MeV bump signal. However, a 5~MeV signal bump could be observed as fake IBD events in IsoDAR@Yemilab from enhancements associated with the $^{\mathrm{13}}\mathrm{C}(\overline{\nu},\overline{\nu}'n)^{\mathrm{12}}\mathrm{C}^{*}$ reaction in the liquid scintillator detector, based on either minimal or non-minimal new physics scenarios, as suggested in Ref.~\cite{Berryman:2018jxt}. IsoDAR@Yemilab has an advantage to test this thanks to the high flux of $\overline{\nu}_e$ above 9.4~MeV required for this reaction (see Fig.~\ref{flux}). In addition, it is interesting to note that the characteristic shape of the low mass
mediator induced IBD event spectrum, shown in Fig.~\ref{x_nunu_spectra}, is similar to the reactor bump for some $X$ masses. In any case, the IsoDAR result will provide an important clue to the source of the reactor bump.


\section{Sensitivity to new physics via \boldmath$\bar{\nu}_e-e^-$ scattering}

In addition to the bounty of IBD events described in the previous section, the IsoDAR@Yemilab configuration will provide about 6980~detected ES events with $E_{\mathrm{vis}}>3$~MeV in a 1.16~kton fiducial volume (6.0~m radius, 12~m height) with a 4~year livetime. Note that this fiducial volume is smaller than the double-flash IBD analysis one described above, due to the single-flash nature of ES events and correspondingly higher background. The $E_{\mathrm{vis}}>3$~MeV requirement is necessary to mitigate rapidly rising radiogenic-induced gammas below this energy cutoff. The event rate estimate includes a 32\% detection efficiency above 3~MeV. Notably, this sample will be significantly higher than the 2600~ES events expected ($E_{\mathrm{vis}}>$3~MeV) in the IsoDAR@KamLAND configuration studied in Ref.~\cite{toups}, driven by the larger fiducial volume of the Yemilab detector.

In this section, we consider these ES signal events, along with relevant backgrounds and systematic uncertainties, in the context of both searching for new physics via non-standard neutrino interactions (NSI) and measuring the weak mixing angle under a ``no NSI" assumption. However, it is important to note that, beyond searching for NSI using ES, the study of the single-electron signature can provide powerful tests of new physics in other ways, which are not explored here in detail (in favor of analyzing the \textit{singular} measurement/observable, of electron kinetic energy in $\bar{\nu}_e-e^-$ scattering, in the context of the weak mixing angle and NSI). For example, neutrino electromagnetic properties~\cite{Giunti:2008ve,BahaBalantekin:2018ppj} can distort the expected SM cross section as well. If neutrinos have a large magnetic moment, as is the case in some neutrino-mass-generation scenarios~\cite{Voloshin:1987qy,LEURER199081}, they would produce observable signatures in experiments measuring small electron recoils~\cite{Shoemaker:2018vii,Brdar:2020quo}. 

This section derives largely from the highly analogous study in Ref.~\cite{toups}, which details an ES measurement in the IsoDAR@KamLAND configuration.

The SM's ES differential cross section is given by:
\begin{align}
\frac{d\sigma}{dT} = \frac{2 G_F^2 m_e}{\pi}
\left[ 
g^2_R + g^2_L\left(1 - \frac{T}{E_\nu}\right)^2 - g_R g_L \frac{m_e T}{E^2_\nu}
\right]~,
\label{glgrxs}
\end{align}

where $g_R= {{1}\over{2}}(g_V-g_A)$, $g_L={{1}\over{2}}(g_V+g_A)$, $E_{\nu}$ is the $\bar{\nu}_{e}$ energy, $T$ is
the electron's recoil kinetic energy, $m_e$ is the mass of the electron, and $G_F$ is the Fermi coupling constant.  The coupling constants are expressed at tree
level as:
\begin{equation}
g_L={{1}\over{2}}+\sin^2\theta_W~, ~~g_R=\sin^2\theta_W~.
\end{equation}

An ES measurement as a function of the outgoing electron's energy can therefore provide a measurement of $g_V$ and $g_A$, as well as $\sin^2{\theta_W}$. The weak mixing angle is constrained experimentally in other ways, of course, and the relationship between $\sin^2{\theta_W}$, $G_F$, $M_Z$ and $\alpha$ [$\sin^2 2\theta_W=(4 \pi \alpha)/(\sqrt{2} G_F M_Z^2)$] combined with precision measurements 
at colliders~\cite{Erler:2013xha} and from muon decay~\cite{Webber:2011ns} provide a precise global prediction for $\sin^2{\theta_W}$, assuming that these measurements are fully applicable and transferable for the neutrino case. However, deviations from this expectation may appear due to NSI. Under different assumptions, the ES study described below can be considered in the context of a $\sin^2{\theta_W}$ measurement or in terms of a search for NSI (while setting $\sin^2{\theta_W}$ to a constant or, alternatively, allowing it to deviate based on a precision measurement in a different sector and assuming universal applicability of the parameter), but the experimental observable and backgrounds are identical for the two cases, although the normalization and energy dependence of the signal expectation can change based on the relevant model's parameters (either $\sin^2{\theta_W}$ or the NSI parameters described below). 

For an NSI search, we can parameterize the deviations from the SM prediction as changes to the relevant couplings:

\begin{eqnarray}\label{epsilonxsec}
\hspace{-1cm}
\frac{d\sigma(E_{\nu}, T)}{dT} &=& \frac{2 G_F^2 m_e}{\pi}\left[\left(\tilde
g_R^2+\sum_{\alpha \neq e}
|\epsilon_{\alpha e}^{e R}|^2\right) + \nonumber \right. \\
& &\left(\tilde g_L^2+ \sum_{\alpha \neq e}
|\epsilon_{\alpha e}^{e L}|^2\right)\left(1-{T \over E_{\nu}}\right)^2 -
\nonumber\\ 
& & \left(\tilde g_R \left.\tilde g_L+ \sum_{\alpha \neq e}|\epsilon_{\alpha e}^{e R}||
\epsilon_{\alpha e}^{e L}|\right)m_e \frac{T}{E^2_{\nu}}\right]~,
\end{eqnarray}
where $\tilde g_R= g_R+\epsilon_{e e}^{e R}$ and $\tilde g_L=g_L+\epsilon_{e e}^{e L}$.  

The NSI parameters ($\epsilon_{e\mu}^{eL,eR}$
and $\epsilon_{e\tau}^{eL,eR}$) are associated with flavor-changing neutral currents. We neglect (set to zero) these parameters when considering IsoDAR@Yemilab's sensitivity to NSI, since they are tightly constrained for muon flavor~\cite{Davidson:2003ha} and lepton-flavor-changing experiments, in general. The $\epsilon_{ee}^{eL,eR}$ are called ``non-universal parameters" and the IsoDAR@Yemilab sensitivity quoted below is in terms of these, with the other four set to zero.

\subsubsection{Signal and Background}

\begin{figure*}[h]
\begin{centering}
\centering
\includegraphics[width=17.5cm]{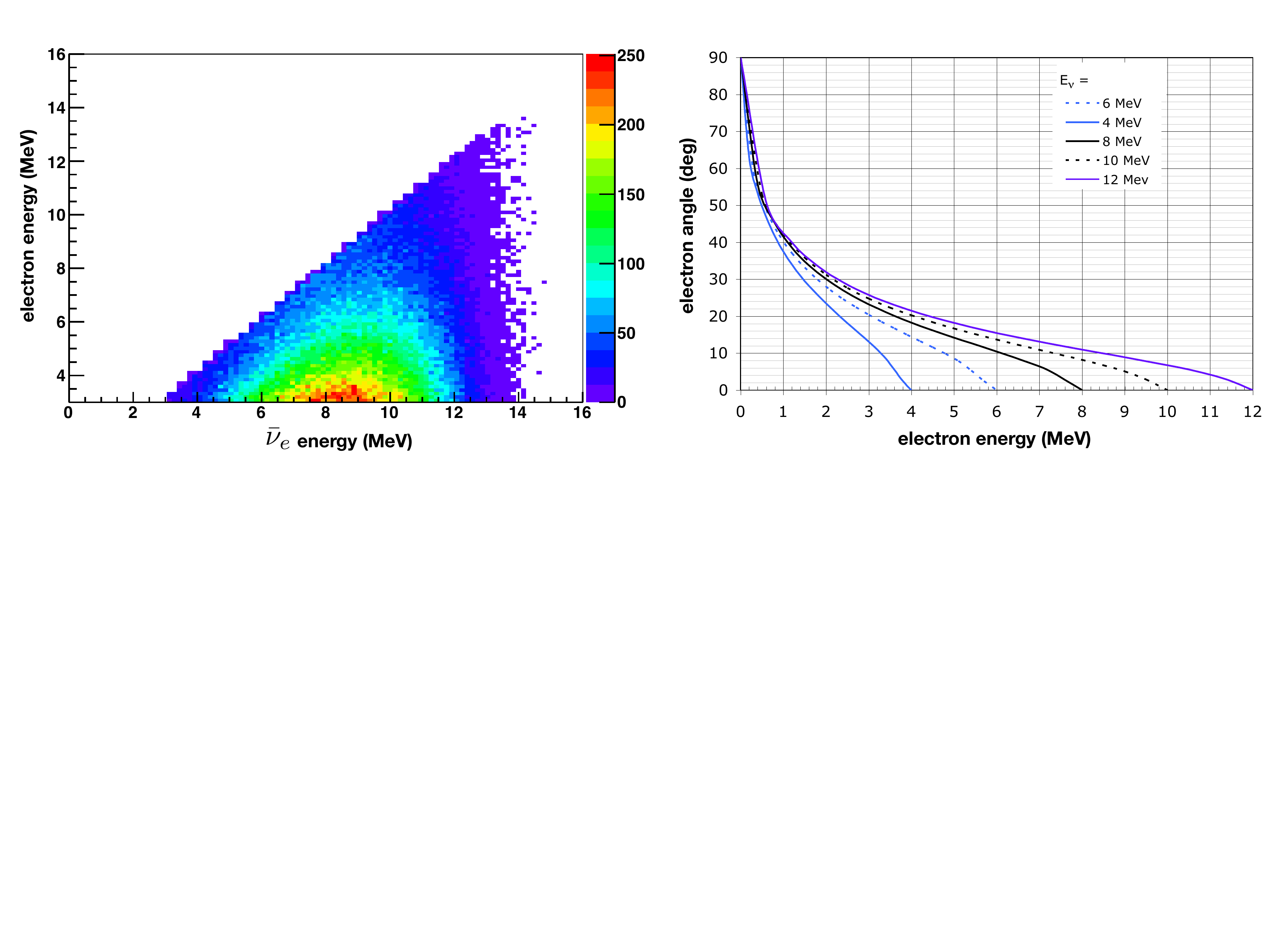}
\vspace{-7cm}
\caption{The kinematics of the outgoing electron in ES ($\bar{\nu}_e + e^- \rightarrow \bar{\nu}_e + e^-$) events in terms of $\bar{\nu}_e$ energy, e$^-$ angle, and e$^-$ kinetic energy. The plot on the left has been produced with the IsoDAR-specific $\bar{\nu}_e$ flux from $^8$Li beta decay.}
\label{angle_stuff}
\end{centering}
\end{figure*}

The signature of an ES interaction ($\bar{\nu}_e + e^- \rightarrow \bar{\nu}_e + e^-$) is simply an outgoing electron, and these events are completely characterized by the recoiling electron's energy ($E_e$) and angle ($\theta _e$), with the recoil energy directly proportional to visible light in the detector, $E_\text{vis}$. The $\bar{\nu}_e$ energy can be reconstructed using these quantities with the following equation: $E_{\bar{\nu}_{e}} = \frac{m_e(E_e - m_e)}{\cos(\theta _e) P_e - E_e + m_e}$. The kinematics associated with ES events generically and in IsoDAR@Yemilab are shown in Fig.~\ref{angle_stuff}. While the angular information is useful for background mitigation, as discussed below, we focus on reconstructing the well-predicted ES $E_{\mathrm{vis}}$ distribution for achieving sensitivity to $\sin^2{\theta_W}$, and consider a detector \textit{without} angular reconstruction abilities as the \textit{default} design. However, we briefly consider an \textit{alternative} scenario in which the detector has directional resolution capabilities below. The $e^-$ energy and vertex resolution assumptions for ES detection are identical to the IBD-induced $e^+$ ones discussed above. These, along with a number of other relevant assumptions, are shown in Table~\ref{assumptions_table_ES} and presented in detail below. 

\begin{table*}[h]
\textbf{ES analysis assumptions}  \\ 
      \begin{tabular}{|c|c|} \hline 
IsoDAR@Yemilab fiducial mass  &  1.16 kton  \\ \hline
IsoDAR@Yemilab fiducial size (radius, height)  &  6.0 m, 12.0 m \\ \hline
e$^-$ energy res.   &  $\sigma(E)=6.4\%/\sqrt{E~\mathrm{(MeV)}}$   \\ \hline
e$^-$ energy res. @ 8~MeV  &  2.3\%  \\ \hline
IBD bkgnd. rejection efficiency~\cite{toups} & $99.75 \pm 0.02\%$ \\ \hline
IBD efficiency unc. (for flux normalization)~\cite{Gando}  & 0.7\%  \\ \hline
e$^{\pm}$ detection eff. for $E_{vis}>3$~MeV (signal and bkgnd.)~\cite{toups}  &  32\%\\ \hline \hline
Total detected $\bar{\nu}_e$-e for $E_{vis}>3$~MeV (w/ 32\% eff.) &  6980  \\ \hline
\end{tabular}
\caption{The assumptions relevant for the IsoDAR@Yemilab $\bar{\nu}_e$ elastic scattering ($\bar{\nu}_e + e^- \rightarrow \bar{\nu}_e + e^-$) analysis.}
\label{assumptions_table_ES}
\end{table*}

Given the single-flash nature of ES events, there are a number of important backgrounds to consider, including mis-reconstructed IBD events from the IsoDAR source, solar $\nu$-induced ES events (the dominant source is $^8$B neutrinos), cosmogenic-spallation-induced isotopes, and radiogenics originating from the liquid, stainless steel, and rock outside of the detector. These backgrounds are discussed below.

The nearly 1.7~~million IBD events expected in 5~years of running with IsoDAR@Yemilab discussed above are useful for an ES-based measurement, in the sense that they constrain the normalization of the $\bar{\nu}_e$ flux (at least, in the absence of oscillations, which is assumed in this section) at the level of the uncertainty in the IBD efficiency of 0.7\%~\cite{Gando}, noting that the IBD cross section error is subdominant at 0.2\% and the statistical uncertainty on the IBD measurement will be at the 0.1\% level. However, the IBD events, nominally \textit{double}-coincident with a prompt $e^+$ followed by a neutron capture signal on either $^1$H (2.2~MeV $\gamma$) or $^{12}$C (4.95~MeV $\gamma$), can pose a background for the ES measurement as well. IBD events are nominally identified by the detection of this delayed neutron capture, with a time constant of $207.5\pm2.8~\mu$s for capture on hydrogen after the $\bar{\nu}_e$ interaction~\cite{Gando}. However, IBD events can be misidentified as \textit{single}/prompt-only $\bar{\nu}_e - e^-$ ES-like events if the neutron is not detected. While IBD identification (or, background rejection in this case) is highly effective, the $\sim50\times$ higher IBD event rate compared to the ES event rate, means that this background is important.
Following the IsoDAR@KamLAND study in Ref.~\cite{toups}, the IBD background rejection efficiency for IsoDAR@Yemilab is conservatively estimated as $99.75 \pm 0.02\%$. This inefficiency is dominated by events which simultaneously produce a reconstructed IBD prompt $e^+$ vertex in the fiducial volume and a neutron capture outside both the fiducial volume and a larger ``neutron capture volume". This estimate, originally based on the KamLAND spherical geometry (5.0~m radius fiducial volume and 6.0~m radius neutron capture volume), is conservative since the IsoDAR@Yemilab cylindrical geometry and distance between the fiducial and neutron capture volumes defined here (6.0~m radius and 12~m height fiducial volume; and 7.5~m radius and 15~m height neutron capture volume) will provide comparatively higher neutron detection efficiency. Ultimately, an AmBe source delayed neutron capture calibration campaign at Yemilab will provide precision estimates of the IBD rejection efficiency and uncertainty for this analysis. 

Neutrino-electron ($\nu_x + e^- \rightarrow \nu_x + e^-$) elastic scattering events from $^8$B solar $\nu_x$ (where $x=e,~\mu,\mathrm{or}~\tau$) also present a background to the $\bar{\nu}_e - e^-$ ES measurement described here.  The relevant solar neutrino flux~\cite{Super-Kamiokande:2010tar} and interaction rate ($4.10\pm0.11$ events/kton$\cdot$day), and energy dependence are described in Ref.~\cite{toups} and are included here, scaled appropriately based on the fiducial volume difference between IsoDAR@KamLAND to the IsoDAR@Yemilab. Radiogenic-induced gammas from the stainless steel containment vessel and rock surrounding the detector also represent backgrounds for the ES measurement. We utilize the \textsc{Geant4}-based background prediction from Ref.~\cite{toups}, scaled according to the geometric (spherical vs. cylindrical) and volume differences between IsoDAR@KamLAND and IsoDAR@Yemilab. Radiogenic daughters from the $^{238}$U and $^{232}$Th decay chains inside of the liquid scintillator can also produce backgrounds. For $E_{\mathrm{vis}}>3$~MeV, the beta decay of $^{208}$Tl ($\tau=3.05$~minutes, $Q$=5.0~MeV) from the $^{232}$Th chain is the main concern. Ref.~\cite{toups} assumed $^{232}$Th contamination at the level of the low-background phase of KamLAND [$(1.12\pm0.21) \cdot 10^{-17}$~g/g]. However, the Yemilab detector expects to achieve significantly better liquid scintillator purity, with $^{232}$Th contamination at the level of $<5.7 \cdot 10^{-19}$~g/g. Here, we conservatively assume that the contamination in the Yemilab detector is equal to this upper-limit.

Another set of backgrounds is due to cosmogenic-induced spallation and the resulting light isotope production, which can produce signal-like events with $E_{\mathrm{vis}}>3$~MeV, including $^8$B ($\beta^+$ decay; $\tau=1.11$~s, $Q$=18~MeV), $^8$Li ($\beta^-$ decay; $\tau=1.21$~s, $Q$=16~MeV), and $^{11}$Be ($\beta^-$ decay; $\tau=19.9$~s, $Q$=11.5~MeV)~\cite{TILLEY2004155,Ajzenberg-Selove:1990fsm}. These backgrounds and their removal, largely based on a cosmic muon veto, are detailed in Ref.~\cite{KamLAND:2009zwo} in the context of IsoDAR@KamLAND, and we apply them here as well, scaled appropriately to the different volumes, noting that the depth-equivalent of each detector is the same (2700~m.w.e.). Again, we expect this estimate to be conservative given the larger distance between the Yemilab detector's fiducial volume's outer edge and the outer edge of the veto region (minimum distance=5.0~m) as compared to KamLAND (4.0~m).

\begin{figure}[h!]
\begin{centering}
\centering
\includegraphics[width=8.5cm]{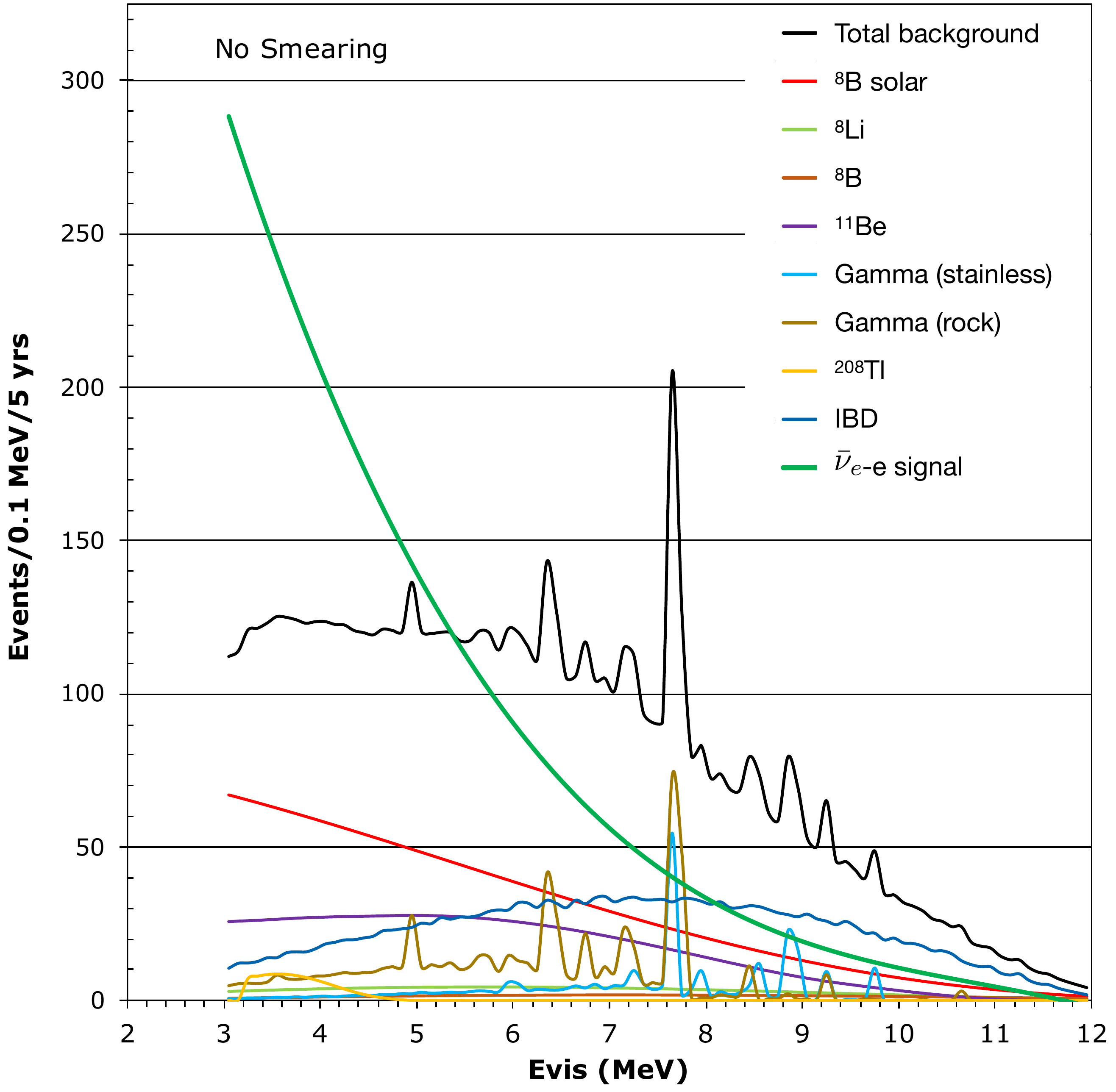}
\caption{The ES signal and background event rates expected in 5~years of running IsoDAR@Yemilab. No energy smearing has been applied to the distributions.}
\label{yemilab_ES_event_Rates}
\end{centering}
\end{figure}

The default scenario (no angular reconstruction capability) signal and background expectations for the ES analysis are summarized in Table~\ref{yemilab_ES_event_Rates_table} and their energy spectra can be seen in Fig.~\ref{yemilab_ES_event_Rates}. After including fiducial volume and veto cuts, the signal event rate detection is 32\% efficient overall above $E_{\mathrm{vis}}>3$~MeV (and 0\% efficient below). The fiducial volume cuts are used to mitigate the radiogenic backgrounds from the stainless steel detector vessel and rock surrounding the detector, and cosmogenic-induced radioactive isotopes. A smaller fiducial volume also limits events featuring IBD-induced neutron escape, and thus IBD misidentification background. The veto cuts are used to reject cosmogenics, and are based on the KamLAND muon veto selection described at Ref.~\cite{KamLAND:2011fld}. In summary, the dominant backgrounds are from solar neutrinos and misidentified IBD events. Solar-induced ES-like events are particularly significant at lower energies, approaching $E_{\mathrm{vis}}>3$~MeV, where signal is rapidly rising. In contrast to the IsoDAR@KamLAND expectation detailed in Ref.~\cite{toups}, $^{208}$Tl is a much smaller background due to the $>20\times$ higher radiopurity envisioned in the Yemilab liquid scintillator.

\subsubsection{Angular reconstruction capability}
\begin{figure}[h]
\begin{centering}
\centering
\includegraphics[width=8.5cm]{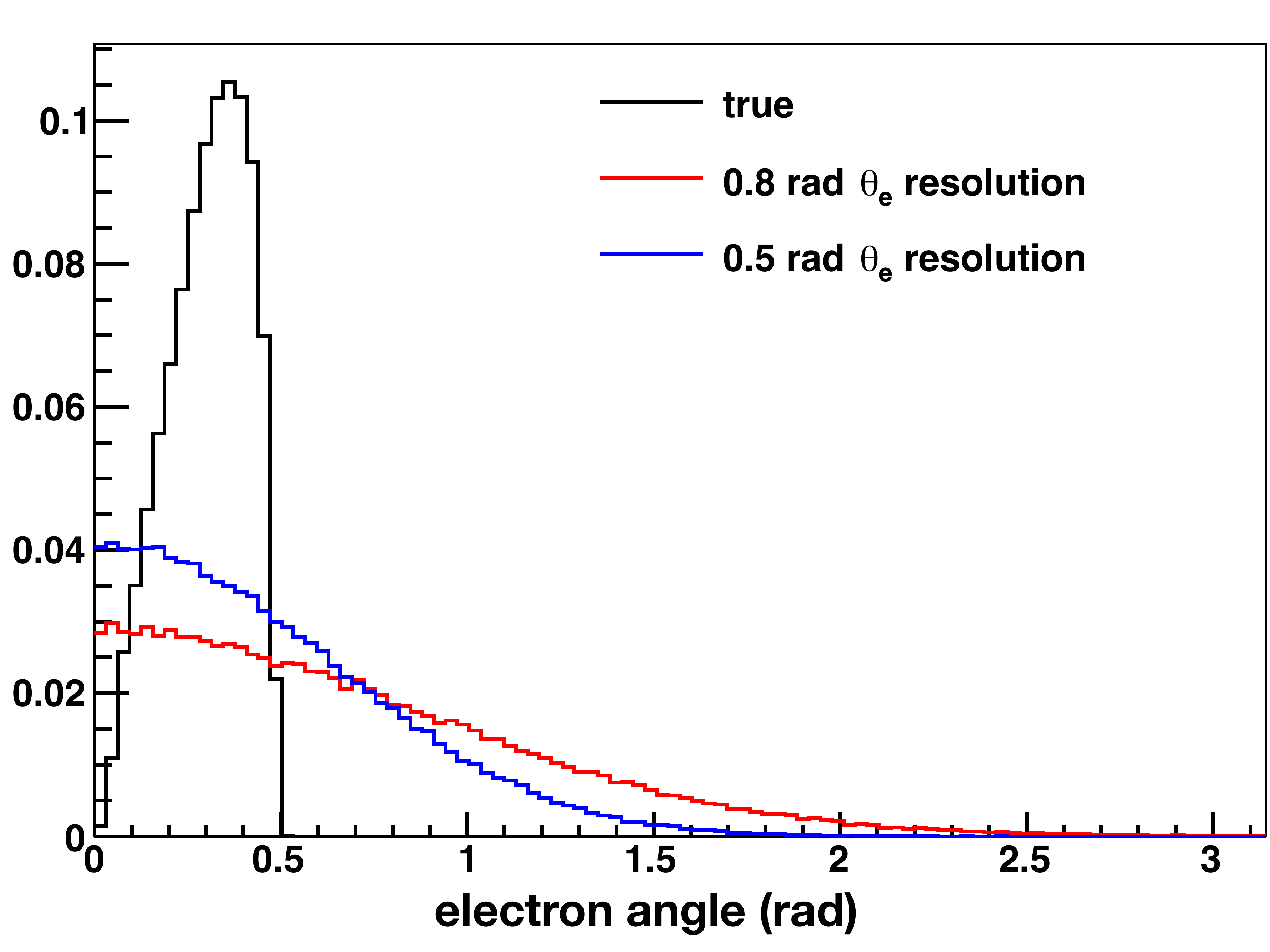}
\caption{The IsoDAR $\bar{\nu}_e$ ES-induced true electron angle and reconstructed angle under two angular resolution assumptions, reasonably consistent with what is expected with the use of water-based liquid scintillator.}
\label{angle_stuff_2}
\end{centering}
\end{figure}

There is a possibility that the Yemilab detector could have angular reconstruction capabilities via some combination of fast-timing photosensors and/or Cerenkov-sensitive photosensors and/or water-based (or dilute) liquid scintillator~\cite{Seo:2019dpr}. Along with allowing $\overline{\nu}_e$ energy reconstruction, the additional ability to reconstruct the direction of the electron in ES events would be a powerful way to mitigate background since the signal electron is very forward (see Fig.~\ref{angle_stuff_2}) and can be pointed back to the IsoDAR target-sleeve $\bar{\nu}_e$ source.  Background events from IBD-induced positrons from the IsoDAR target-sleeve can be considered isotropic.  Backgrounds from solar neutrino ES events, however, are not isotropic; they will generally point directly away from the Sun.  Therefore, solar neutrino ES events will be most likely to pass an angular selection cut only during the parts of the day and year where the Sun is behind the target-to-detector line. The IsoDAR@Yemilab beam-to-detector orientation will be roughly east-to-west aligned, which will mean data taken at sunrise will have significantly more solar neutrino backgrounds than data taken at other times of the day.

 Notably, while these detector ``improvements" would enable electron angular reconstruction, it is also possible that energy and vertex resolution could also be affected, perhaps negatively, by these significant detector changes.  For simplicity, we ignore this possibility and maintain the energy and vertex resolutions assumed above when calculating the weak mixing angle sensitivity under this alternative scenario featuring angular reconstruction. The true electron angle and reconstructed angle under two angular resolution assumptions in ES events are shown in Fig.~\ref{angle_stuff_2}. With an angular resolution of 0.5~rad, reasonably consistent with what might be expected with the use of water-based liquid scintillator, and a simplistic signal selection criterion of $\theta_e<1$~rad, the ES analysis would be 90\% efficient in selecting signal while reducing the isotropic and solar neutrino backgrounds by a factor of 3. This alternative scenario is considered below when calculating IsoDAR@Yemilab's sensitivity to $\sin^2{\theta_W}$ and NSI.

\begin{table}
\begin{center} 
      \begin{tabular}{|c|c|} \hline
Event type  & Counts   \\ 
  & ($E_{\mathrm{vis}}\in3-12$~MeV)/ 5 years  \\ \hline \hline
        $^8$B solar $\nu$ & 2531 \\ 
        $^8$Li spallation& 270 \\ 
        $^8$B spallation & 121  \\ 
        $^{11}$Be spallation& 1393  \\ 
        $\gamma$ rock  & 683 \\ 
        $\gamma$ stainless  & 291\\ 
        $^{208}$Tl & 84  \\
        IBD& 2013  \\ \hline
        Total background & 7387  \\ \hline
        $\bar{\nu}_e$-$e^-$ signal  &  6977 \\ \hline
      \end{tabular}
      \caption{A summary of the ES signal and background events for 5 years runtime (4~years livetime) in Fig.~\ref{yemilab_ES_event_Rates}.}
      \label{yemilab_ES_event_Rates_table}
\end{center}
\end{table}

\subsubsection{Analysis}
As discussed above, the ES differential cross section depends on $\sin^2{\theta_W}$, $g_V$, $g_A$, and NSI parameters. Here, we quote the sensitivity to (1) $\sin^2{\theta_W}$, assuming no NSI, and (2) NSI parameters, while setting $\sin^2{\theta_W}$ to be a constant. For determining the sensitivity to $\sin^2{\theta_W}$, we perform a fit to the ES~signal+background, while relying on expected \textit{in-situ} measurements to provide the normalization of both the antineutrino flux and the beam-on (IBD) background, as well as beam-off measurements (statistical-error only) of steady-state backgrounds. Based on the expected IBD inefficiency discussed above, the misidentified IBD background is estimated as $(0.25\pm0.02)\%$ of the entire IBD event collection.

We estimate the measurement sensitivity of IsoDAR@Yemilab following the methodology outlined in Ref.~\cite{toups}. Signal and background events are binned together by $E_\text{vis}$ in 0.5~MeV bins from 3~MeV to 14~MeV. To estimate the uncertainty achievable on a measurement of $\sin^2{\theta_W}$, we perform a fit using a  $\Delta \chi^2 = \chi^2\text{(fit)} - \chi^2\text{(best fit)}$ statistic. In determining sensitivity, we assume the best fit corresponds to the signal, with $\sin^2\theta _W ^ 0 = 0.238$, and the backgrounds presented above. 

Following Ref.~\cite{toups}, we write $\chi^2$ in the following way. For the $i$th bin in terms of $E_\text{vis}$, we let $ES$ be the number of elastic scattering events at $\sin^2\theta _W = s$. We also define $B^\text{on}_i$ as beam-on backgrounds (IBD) and $B^\text{off}_i$ as non-beam backgrounds (solar, radiogenic, cosmogenic) for the $i$th bin. The number of events in this bin is \begin{equation}
    N_i(s) = ES_i (s) + B^\text{on}_i + B^\text{off}_i
\end{equation}

With $s_0=\sin^2\theta _W ^0$ and $s_f=\sin^2\theta _W ^\text{fit}$, we have:
\begin{multline}
\chi^2(s_f) =\\ \sum_{i}\frac{\left(N_i(s_0) - (N_i(s_f) + \alpha * ES_i(s_f) + \beta * B_i^\text{on})\right)^2}{(N_i(s_0) + B_i^\text{off})} \\ +  \left(\frac{\alpha}{\sigma_\alpha}\right)^2 +\left(\frac{\beta}{\sigma_\beta}\right)^2
\label{chisq}
\end{multline}

The normalization uncertainties for the ES signal and the IBD misidentification background events are included using the pull parameters $\alpha$ and $\beta$, respectively. $\alpha$ is constrained by the uncertainty in the IBD efficiency as $\sigma_\alpha = 0.7\%$. $\beta$ is the uncertainty in the misidentified IBD background, determined above to be 0.25\% of the IBD sample, and we estimate a conservative $\sigma_\beta = 0.02/0.25$ uncertainty on this value. Statistical uncertainties on the beam-off backgrounds are included in the fit as well. 

\begin{figure}[h!]
\begin{centering}
\centering
\includegraphics[width=8.6cm]{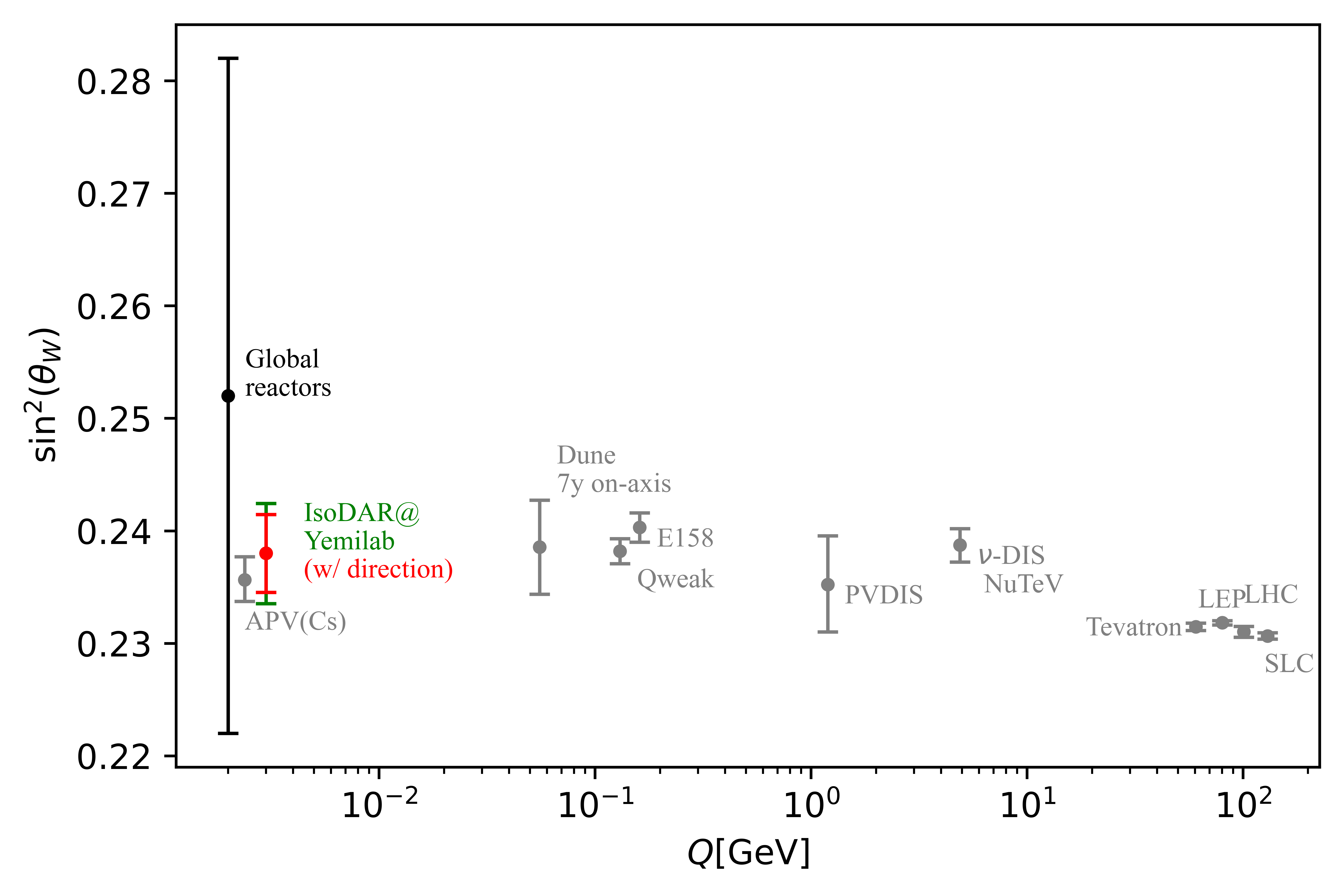}
\caption{IsoDAR@Yemilab's sensitivity to $\sin^2{\theta_W}$ in comparison to past and future (DUNE~\cite{dune_weakmixingangle,dune_nd}) experiments, and a global reactor-antineutrino analysis~\cite{global_reactor}. Aspects of this figure are adapted from Ref.~\cite{dune_weakmixingangle}.}
\label{mixing_angle}
\end{centering}
\end{figure}

\begin{figure}[h!]
\begin{centering}
\centering
\includegraphics[width=8.7cm]{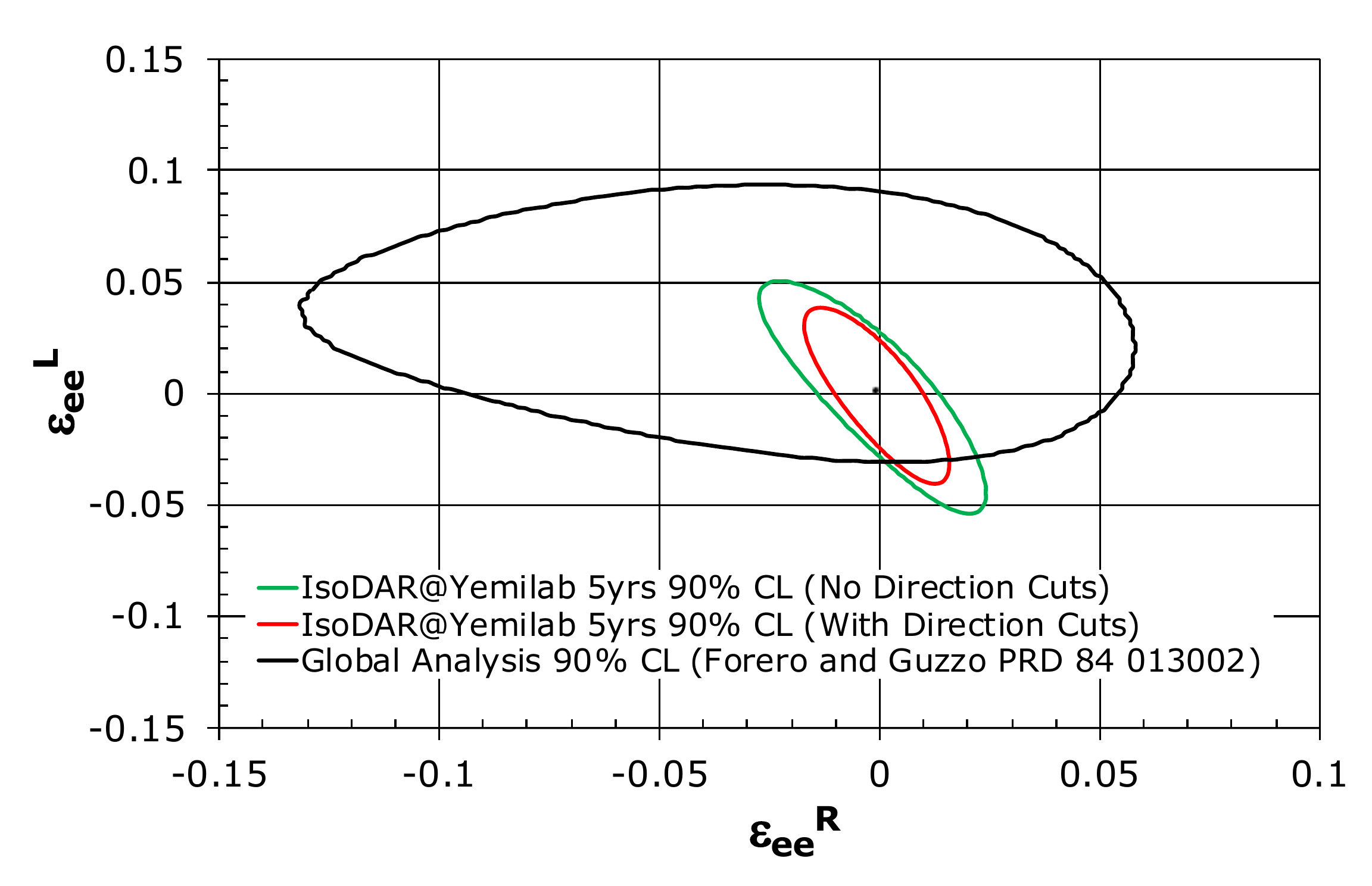}
\caption{The expected achievable sensitivity for IsoDAR@Yemilab's 5-year runtime ES measurement (with and without directional reconstruction capabilities), in terms of the NSI parameters $\epsilon_{e e}^{eLR}$ (near $\epsilon_{e e}^{eLR}\sim0$, noting that there is a four-fold degeneracy of these values, see Eq.~\ref{epsilonxsec}). Also shown is a global fit to these parameters, based on Ref.~\cite{guzzo}.}
\label{yemilab_nsi_sensitivity}
\end{centering}
\end{figure}

The fitting results show that the IsoDAR@Yemilab experiment can expect a $\delta \sin^2{\theta_W}$ sensitivity of 0.0045 (1.9\% measurement), using rate and energy-shape information and including statistical and systematic uncertainties, in 5~years of running. As can be seen in Fig.~\ref{mixing_angle}, this sensitivity would improve upon the current global reactor measurement of $\sin^2\theta_W=0.252\pm0.030$~\cite{global_reactor} by nearly an order of magnitude. This can also be compared to that which is expected in the IsoDAR@KamLAND configuration of a 3.2\% $\sin^2{\theta_W}$ measurement (rate+shape)~\cite{toups}. In the alternative scenario considered, with the ability to reconstruct the electron angle, the IsoDAR@Yemilab sensitivity in terms of $\delta \sin^2{\theta_W}$ would improve to 0.0035 (1.5\% measurement). In terms of an NSI search, after fixing $\sin^2{\theta_W}=0.238$, the achievable sensitivity to the NSI non-universal parameters $\epsilon_{ee}^{eL,eR}$ in both the w/ and w/o directional reconstruction capability scenarios is shown in Fig.~\ref{yemilab_nsi_sensitivity}. As can be seen, the IsoDAR@Yemilab experiment would provide greatly improved sensitivity to NSI as compared to a current global fit~\cite{guzzo}.

\section{Conclusions}
The IsoDAR electron antineutrino source combined with a kiloton-scale detector at Yemilab would provide unprecedented sensitivity to new physics via (1) a search for short-baseline oscillations, including initial-state wavepacket effects, and the ability to trace the $L/E$ wave with a collection of $1.7\cdot10^6$ $\bar{\nu}_e$-induced IBD events; (2) a search for \textit{other} unexpected deviations in this IBD sample (e.g. a bump hunt), which are, for example, motivated by theory models involving light mass mediators, and experiment, including the X17 particle and the 5~MeV reactor bump anomalies; and (3) a precision measurement of $\bar{\nu}_e$-induced electron scattering events as an electroweak probe and search for non-standard neutrino interactions. The latter measurement would be significantly enhanced by the detector's potential capability to reconstruct the direction of signal electrons. These physics studies would greatly improve upon existing measurements, in particular, at a level approaching an order of magnitude in both sterile-oscillation and weak-mixing-angle/NSI sensitivity.

\section{Acknowledgements}
We thank P. Denton and M. Hostert for useful discussions. CAA is supported by the Faculty of Arts and Sciences of Harvard University, and the Alfred P. Sloan Foundation. JMC and DW are supported by National Science Foundation award \#1912764. YDK and SHS are supported by IBS-R016-D1. JS is supported by National Science Foundation award \#2012897. MHS is supported by National Science Foundation award \#PHY-2013070. 

\bibliography{main}

\end{document}